\documentclass[12pt]{iopart}

 \usepackage{hyperref} 
\usepackage{graphicx}
\usepackage{epstopdf}

\usepackage{tikz}
\usetikzlibrary{arrows}
\usetikzlibrary{positioning}
\usetikzlibrary{decorations.text}
\usetikzlibrary{decorations.pathmorphing}

 \usepackage{amssymb}

\usepackage{bm}
\begin{document}

\title[Towards a general(ized) shear thickening rheology of wet granular materials under small pressure]{Towards a general(ized) shear thickening rheology of wet granular materials under small pressure}

\author{Sudeshna Roy, Stefan Luding and Thomas Weinhart}

\address{Multi Scale Mechanics (MSM), MESA+, CTW, University of Twente, PO Box 217, 7500 AE Enschede, The Netherlands}
\ead{s.roy@utwente.nl}
\vspace{10pt}
\begin{indented}
\item[]14th August 2016
\end{indented}

\begin{abstract}
We study the rheology of dry and wet granular materials in the steady quasistatic regime using the Discrete Element Method (DEM) in a split-bottom ring shear cell with focus on the macroscopic friction. The aim of our study is to understand the local rheology of bulk flow at various positions in the shear band, where the system is in critical state. The general(ized) rheology has four dimensionless control parameters that relate the time scales of five significant phenomena, namely, the time scales related to confining pressure $t_p$, shear rate $t_{\dot{\gamma}}$, particle stiffness $t_k$, gravity $t_g$ and cohesion $t_c$, respectively. We show that those phenomena collectively contribute to the rheology as multiplicative correction functions. 
\par
While $t_{\dot{\gamma}}$ is large and thus of little importance for most of the data studied, it can increase the friction of flow in critical state, where the shear gradients are high. $t_g$ and $t_k$ are comparable to $t_p$ in the bulk, but become more or less dominant relative to $t_p$ at the extremes of the free surface and deep inside the bulk, respectively.
\par
 We also measure the effect of strong wet cohesion on the flow rheology, as quantified by decreasing $t_c$. Furthermore, the proposed rheological model predicts well the shear thinning behavior both in the bulk and near the free surface; shear thinning develops to shear thickening near the free surface with increasing cohesion.
 
\end{abstract}

%
%
%
%
%

\section{Introduction}

The ability to predict a material's viscosity gives manufacturers an important product dimension. Knowledge on material's rheological characteristics is important in predicting the pourability, density and ease with which it may be handled, processed or used. The interrelation between rheology and other product dimensions often makes the measurement of viscosity the most sensitive or convenient way of detecting changes in flow properties. A frequent reason for the measurement of rheological properties can be found in the area of quality control, where raw materials must be consistent from batch to batch. For this purpose, flow behavior is an indirect measure of product consistency and quality.
\par
Most studies on cohesive materials in granular physics focus on dry granular materials or powders and their flow \cite{imole2014experiments,singh2014effect}. However, wet granular materials are ubiquitous in geology and many real-world applications where interstitial liquid is present between the grains. Many studies have applied the $\mu\left(I\right)$-rheology to flows of dry materials at varying inertial numbers $I$ \cite{singh2015role,thornton2012frictional,thornton2013multi,weinhart2013coarse,vescovi2016merging}. Studies of wet granular rheology include flow of dense non-Brownian suspensions \cite{bonnoit2010mesoscopic,huang2007viscosity,huang2005flow,lemaitre2009dry}. Here, we study partially wetted system of granular materials, in particular the pendular regime, which is also covered in many studies \cite{roy2015microtextendashmacro,schwarze2013rheology,willett2000capillary}. One of the important aspects of partially wetted granular shear flows is the dependence of shear stress on the cohesive forces for wet materials. Various experimental and numerical studies show that addition of liquid bridge forces leads to higher yield strength. The yield stress at critical state can be fitted as a linear function of the pressure with the friction coefficient of dry flow ${\mu}_{o}$ as the slope and a finite offset $c$, defined as the steady state cohesion in the limit of zero confining pressure \cite{roy2015microtextendashmacro}. This finite offset $c$ is constant in the high pressure limit. However, very little is known regarding the rheology for granular materials in the low pressure limit.
\par
Depending on the surrounding conditions, granular flows phenomenon are affected by appropriate time scales namely, $t_{p}$: time required for particles to rearrange under certain pressure, $t_{\dot{\gamma}}$: time scale related to strain rate $\dot{\gamma}$, $t_{k}$: related to the contact time between particles, $t_{g}$: elapsed time for a single particle to fall through half its diameter under the influence of gravity and $t_{c}$: time scale for the capillary forces driving the flow are primarily hindered by inertia based on particle density. While various time scales, as related to the ongoing mechanisms in the sheared bulk of the material, can interfere, they also can get decoupled, in the extremes of the local/ global condition, if one time scale gets way smaller in magnitude than the other. A detailed description of this time scales are given in Sec. \ref{timescale}. While $t_{k}$, $t_{g}$ and $t_{c}$ are global, other time scales $t_{\dot{\gamma}}$ and $t_{p}$ depends on local field variables strain rate ${\dot{\gamma}}$ and pressure $p$ respectively. We restrict our studies to the quasi-static regime ($t_{\dot{\gamma}} \gg t_{{p}}$) as the effect of cohesion decreases with increasing inertial number due to the fast decrease in coordination number \cite{berger2016scaling}. In the present work, we shed light on the rheology of non-cohesive dry as well as cohesive wet granular materials at the small pressure limit, by studying free surface flow. While the inertial number $I$ \cite{koval2009annular}, i.e. the ratio of confining pressure to strain-rate time scales, is used to describe the change in flow rheology from quasi-static to inertial conditions, we look at additional dimensionless numbers that influence the flow behavior. (i) The local compressibility $p^*$, which is the squared ratio of the softness and stress time scales (ii) the inverse relative pressure gradient ${p_g}^*$, which is the squared ratio of gravitational and stress time scales and (iii) the Bond number $Bo$  \cite{bond1937nature} quantifying local cohesion as the squared ratio of stress to wetting time scales are these dimensionless numbers. Additional relevant parameters are not discussed in this study, namely granular temperature or fluidity. All these dimensionless numbers can be related to different time scales or force scales relevant to the granular flow.  
\par
Granular materials display non-Newtonian flow behavior for large enough shear stress while they remain mostly elastic like solids below this yield stress. More precisely, granular materials flow like shear thinning fluid under sufficient stress. When dealing with wet granular materials, a fundamental question is, what is the effect of cohesion on the bulk flow and yield behavior? The second part of this paper is devoted to study of this behavior of granular materials with increasing cohesion. A typical example of interesting rheology of cohesive materials is shown by cement paste. In flow, this is shear thinning at low shear rates and becomes shear thickening at higher shear rates. Microscopically, the shear rate affects the kinetics of the cluster formation where the bonds experience higher stress than average and thereby increasing their persistence. Shear thickening is often observed in concentrated colloidal suspensions due to the formation of jamming clusters resulting from hydrodynamic lubrication forces between particles, also denoted hydroclusters \cite{fall2012shear,wagner2009shear}. Thus shear thickening is related to cluster formation. Highly cohesive wet granular materials have high local Bond number, especially near to the free surface, where the effect of repulsive force is less dominant than the attractive counterpart. This high local formation leads to the formation of local clusters in the system, thereby changing the shear-thinning properties.

\section{Model System}
\subsection{Geometry}
\begin{figure}[!htb]
  \begin{center}
  {%
	\includegraphics[width=0.5\columnwidth]{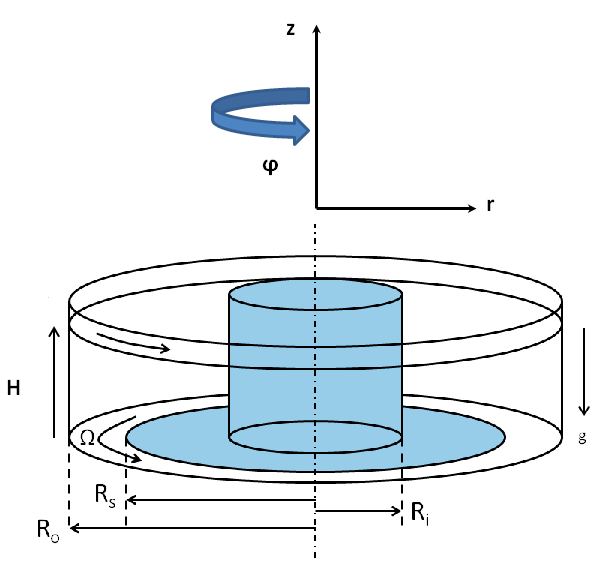}
   }%
  \end{center}
  \caption{Shear cell set-up.}
  \label{fig:setup}
\end{figure}

\textit{Split- Bottom Ring Shear Cell:} We use MercuryDPM \cite{thornton2012modeling,weinhart2012from}, an open-source implementation of the Discrete Particle Method, to simulate a shear cell with annular geometry and a split bottom plate, as shown in Figure \ref{fig:setup}. Some of the earlier studies in similar rotating set-ups include \cite{schollmann1999simulation,wang2012microdynamic,woldhuis2009wide}. The geometry of the system consists of an outer cylinder (outer radius $R_\mathrm{o}$ = 110 mm) rotating around a fixed inner cylinder (inner radius $R_\mathrm{i}$ = 14.7 mm) with a rotation frequency of $\Omega$ = 0.01 revolutions per second. The granular material is confined by gravity between the two concentric cylinders, the bottom plate, and a free top surface. The bottom plate is split at radius $R_\mathrm{s}$ = 85 mm. Due to the split at the bottom, a narrow shear band is formed. It moves inwards and widens towards the flow surface. This set-up thus features a wide shear band away from the bottom and the side walls which is thus free from boundary effects. The filling height ($H$ = 40 mm) is chosen such that the shear band does not reach the inner wall at the free surface. 

\par
In earlier studies \cite{roy2015towards,singh2014effect,singh2015role}, a quarter of this system (${0}^{\circ}$ $\leq$ $\phi$ $\leq$ ${90}^{\circ}$) was simulated using periodic boundary conditions. In order to save computation time, here we simulate only a smaller section of the system (${0}^{\circ}$ $\leq$ $\phi$ $\leq$ ${30}^{\circ}$) with appropriate periodic boundary conditions in the angular coordinate, unless specified otherwise. We have observed no noticeable effect on the macroscopic behavior in comparisons between simulations done with a smaller (${30}^{\circ}$) and a larger (${90}^{\circ}$) opening angle. Note that for very strong attractive forces, agglomeration of particles occur. Then, a higher length scale of the geometry is needed and thus the above statement is not true anymore.
\subsection{Contact model and parameters}

The liquid bridge contact model is based on a combination of an elastic-dissipative linear contact model for the normal repulsive force and a non-linear irreversible liquid bridge model for the non-contact adhesive force as described in \cite{roy2015microtextendashmacro}. The adhesive force is determined by three parameters; surface tension $\gamma$, contact angle $\theta$ which determine the maximum adhesive force and the liquid bridge volume $V_\mathrm{b}$ which determines the maximum interaction distance between the particles at the point of bridge rupture. The contact model parameters and particle properties are as given in Table \ref{constants1}. We have a polydisperse system of glass bead particles with mean diameter ${d}_\mathrm{p} = {\langle d \rangle} = 2.2$ mm and a homogeneous size distribution ($d_\mathrm{min}/d_\mathrm{max} = 1/2$ of width $1 - {\langle d \rangle}^2/{\langle d^2 \rangle} \approx 0.19$).
\begin{table}
\center
\caption{\label{constants1}Table showing the particle properties and constant contact model parameters.}
\footnotesize
\begin{tabular}{@{}lll}
\br
Parameter & Symbol & Value\\
\mr
Sliding friction coefficient & $\mu_p$ & 0.01 \\ 
Normal contact stiffness & $k$ & $120$ N$\,$m$^{-1}$ \\  
Viscous damping coefficient & ${\gamma}_o$ & $0.5\times $10$^{-3}$ kg$\,$s$^{-1}$ \\
Rotation frequency  & $\Omega$ &  $0.01$ s$^{-1}$ \\
Particle density  & $\rho$ &  $2000$ kg$\,$m$^{-3}$ \\
Gravity  & $g$ &  $9.81$ m$\,$s$^{-2}$\\
Mean particle diameter  & $d_\mathrm{p}$ &  $2.2$ mm \\
Contact angle  & $\theta$ &  ${20}^\circ$\\
Liquid bridge volume  & $V_\mathrm{b}$ &  ${75}$ nl \\
\br
\end{tabular}\\
\end{table}
\par
To study the effect of inertia and contact stiffness on the non-cohesive materials rheology, we compare our data for non-cohesive case with data from simulations of \cite{singh2015role} for different gravity as given below:
\begin{equation} \label{g}
g \ \in \ \left\{1.0\right., \ 2.0,  \  5.0, \ 10.0, \
 20.0,  \ \left. 50.0\right\} \  \textnormal{m}\,\textnormal{s}^{-2}
\end{equation}
We also compare the effect of different rotation rates on the rheology for the following rotation rates:
\begin{equation} \label{rotr}
\Omega \ \in \ \left\{0.01\right., \ 0.02,  \  0.04, \ 0.10, \
 0.20, \  0.50,  \  0.75,  \ \left. 1.00\right\} \  \textnormal{s}^{-1}
\end{equation}
\par
The liquid capillary force is estimated as stated in \cite{willett2000capillary}. It is observed in our earlier studies \cite{roy2015microtextendashmacro} that the shear stress $\tau$ for high pressure can be described by a linear function of confining pressure, $p$, as $\tau = {\mu}_{o}p + c$. It was shown that the steady state cohesion $c$ is a linear function of  the surface tension of the liquid $\sigma$ while its dependence on the volume of liquid bridges is defined by a cube root function. The friction coefficient ${\mu}_{o}$ is constant and matches the friction coefficient of dry flows excluding the small pressure limit. In order to see the effect of varying cohesive strength on the macroscopic rheology of wet materials, we vary the intensity of capillary force by varying the surface tension of the liquid $\sigma$, with a constant volume of liquid bridges, ($V_\mathrm{b} = 75$ nl) corresponding to a saturation of $8 \%$, as follows:
\begin{equation} \label{Surf}
\sigma \ \in \ \left\{0.0\right., \ 0.01,  \  0.02, \ 0.04, \
 0.06,  \ 0.10,  \   0.20,   \   0.30,  \  0.40, \ \left. 0.50\right\} \  \mathrm{N}\,\mathrm{m}^{-1}
\end{equation}

\begin{figure}[!htb]
		
	\includegraphics[width=0.5\columnwidth]{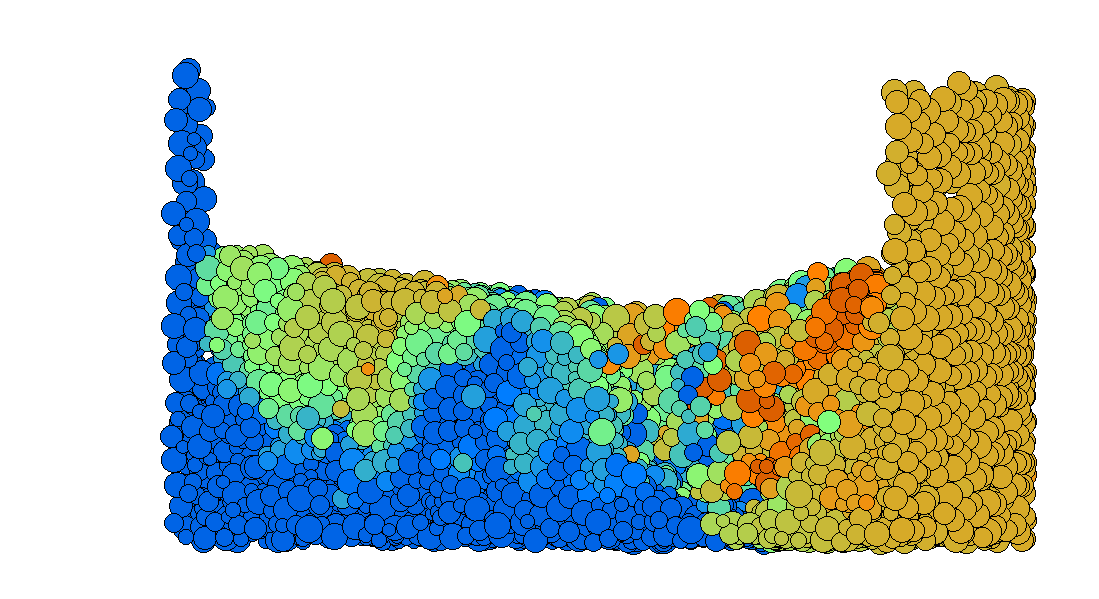}
\hfill
	\includegraphics[width=0.5\columnwidth]{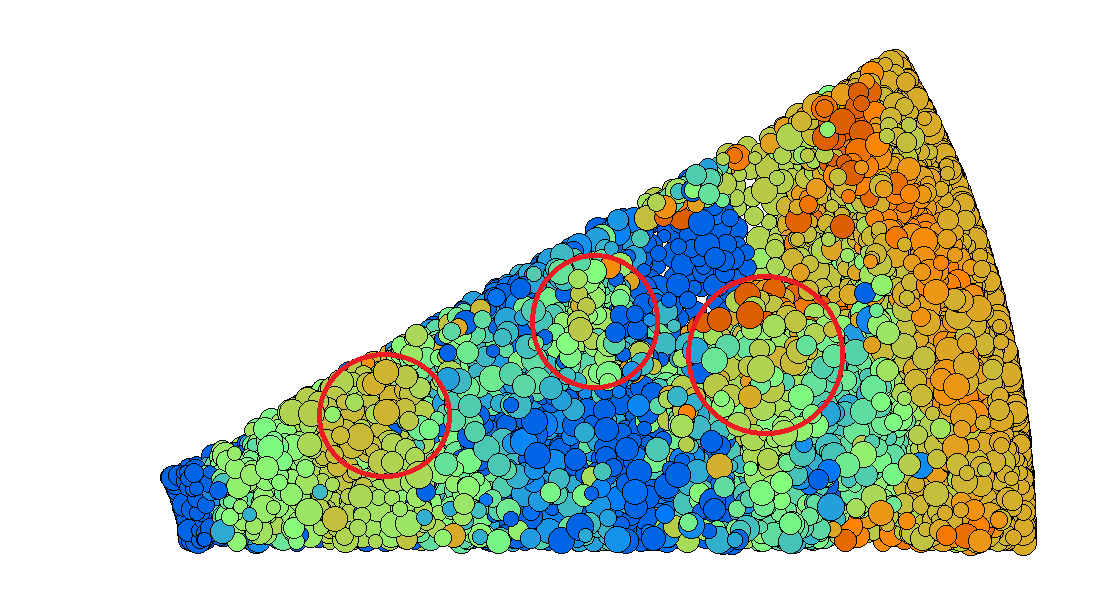}

  \caption{Cluster formation (shown by red circles) for highly cohesive materials ($\sigma = 0.70\,$N$\,$m$^{-1}$) a) front view and b) top view.}
  \label{fig:cluster}
\end{figure}

The first case, $\sigma = 0.0\,$N$\,$m$^{-1}$, represents the case of dry materials without cohesion, whereas $\sigma = 0.50\,$N$\,$m$^{-1}$ corresponds to the surface tension of a mercury-air interface. For $\sigma > 0.50\,$N$\,$m$^{-1}$, smooth, axisymmetric shear band formation is not observed and the materials agglomerate to form clusters as shown in figure \ref{fig:cluster}, for our particle size and density. Hence, $\sigma$ is limited to maximum of $ 0.50\,$N$\,$m$^{-1}$. 
\subsection{Averaging methodology} \label{Averaging}To extract the macroscopic properties, we use the spatial coarse-graining approach detailed in \cite{luding2008constitutive,luding2008effect,luding2011critical}. The averaging is performed over a grid of $47$-by-$47$ toroidal volumes, over many snapshots of time assuming rotational invariance in the tangential $\phi$-direction. The averaging procedure for a three-dimensional system is explained in \cite{luding2008constitutive,luding2011critical}. This spatial coarse-graining method was used earlier in \cite{luding2011critical,roy2015towards,singh2015contact,singh2014effect,singh2015role}. We do the temporal averaging of non-cohesive simulations over a larger time window from $30\,$s to $440\,$s with $2764$ snapshots to ensure the rheological models with enhanced quality data. All the other simulations are run for $200\,$s and temporal averaging is done when the flow is in steady state, between $80\,$s to $200\,$s with $747$ snapshots, thereby disregarding the transient behavior at the onset of the shear. In the critical state, the shear band  is identified by the region having strain rates higher than $80 \%$ of the maximum strain rate at the corresponding height. Most of the analysis explained in the later sections are done from this critical state data at the center of the shear band.
 \subsubsection{Macroscopic quantities}
 The general definitions of macroscopic quantities including stress and strain rate tensors are included in \cite{singh2015role}. Here, we define the derived macroscopic quantities such as the friction coefficient and the apparent viscosity which are the major subjects of our study.
 \par
 The local macroscopic friction coefficient is defined as the ratio of shear to normal stress and is defined as ${\mu} = {\tau}/p$.
 \par
 The magnitude of strain rate tensor in cylindrical polar coordinates is simplified, assuming $u_r = 0$ and $u_z = 0$:
\begin{equation} \label{s_r}
\dot{\gamma} = \frac{1}{2}\sqrt{{\left(\frac{\partial u_{\phi}}{\partial r} - \frac{u_{\phi}}{r}\right)}^2 + {\left(\frac{\partial u_{\phi}}{\partial z}\right)}^2}
\end{equation}
\par
 The apparent shear viscosity is given by the ratio of the shear stress and strain rate as:
\begin{equation} \label{viscosity}
{\eta} = \frac{{{\tau}}}{{\dot{\gamma}}} = \frac{{{\mu}}p}{{\dot{\gamma}}}~,
\end{equation}
where ${\dot{\gamma}}$ is the strain rate. 
\subsection{Critical state}
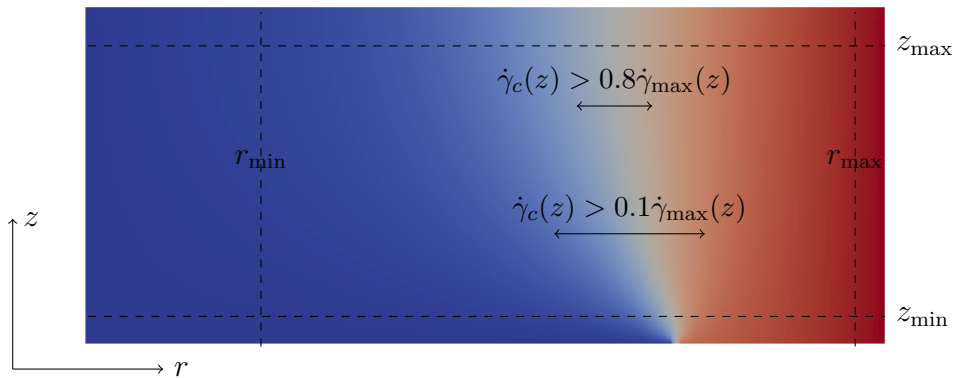
\begin{figure}
\centering
\begin{tikzpicture}
    \node[anchor=south west,inner sep=0] at (-1,-2) {\includegraphics[width=0.8\textwidth]{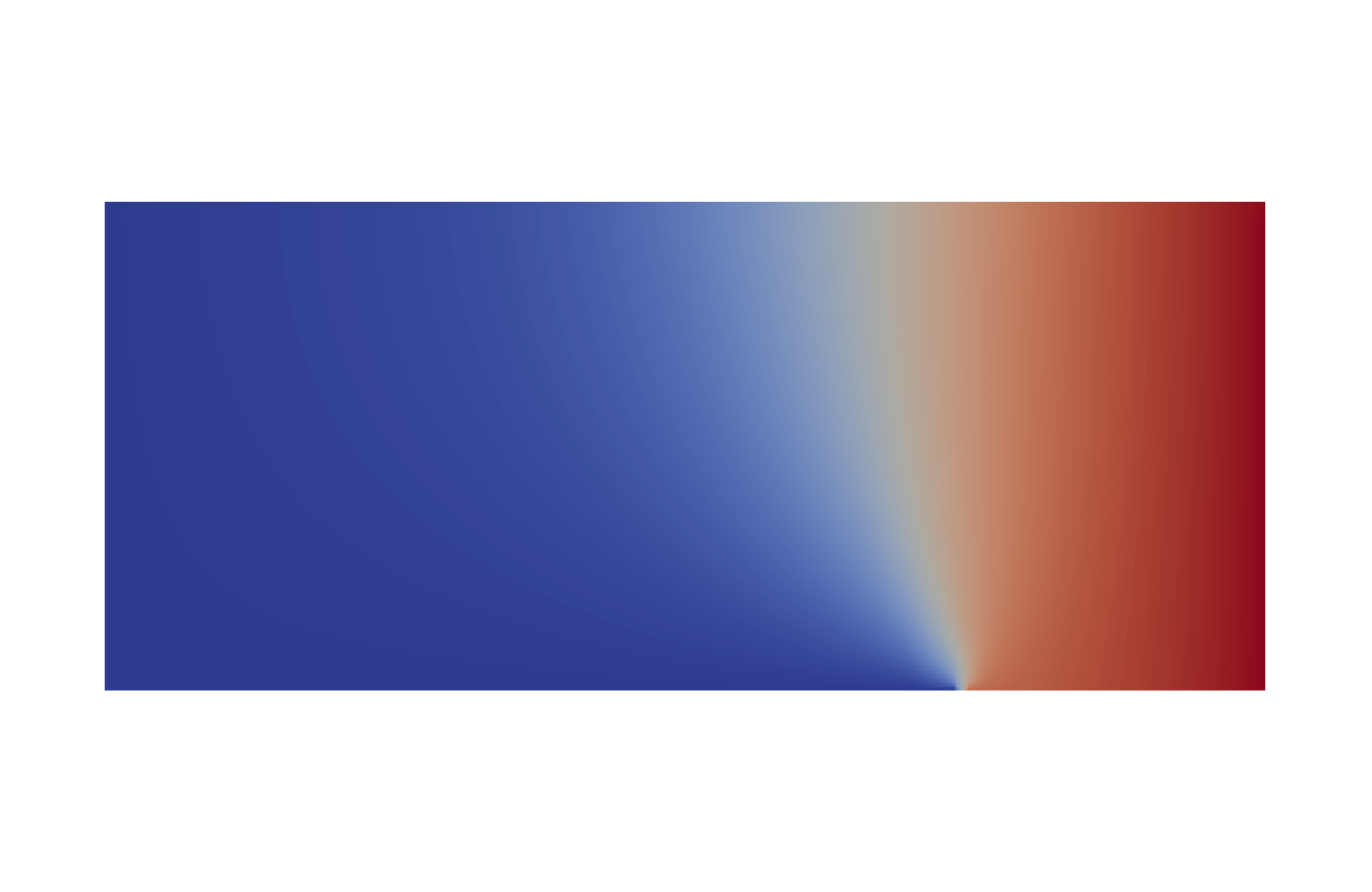}};
    \draw[draw=black,dashed, -stealth, -] (0, 3.8) -- (10.6, 3.8) node[above,right] {$z_{\mathrm{max}}$} ;
    \draw[draw=black,dashed, -stealth, -] (0, 0.2) -- (10.6, 0.2) node[above,right] {$z_{\mathrm{min}}$} ;
    \draw[draw=black,dashed, -stealth, -] (2.3, -0.20) -- (2.3, 4.25) node[above,midway] {$r_{\mathrm{min}}$} ;
        \draw[draw=black,dashed, -stealth, -] (10.2, -0.20) -- (10.2, 4.25) node[above,midway] {$r_{\mathrm{max}}$} ;
\draw[draw=black,solid, -stealth, <->] (6.5, 3) -- (7.5, 3) node[above,midway] { \small ${\dot{\gamma}}_c(z) > 0.8{\dot{\gamma}}_{\mathrm{max}}(z)$} ;

\draw[draw=black,solid, -stealth, <->] (6.2, 1.3) -- (8.2, 1.3)node[above,midway] {\small ${\dot{\gamma}}_c(z) > 0.1{\dot{\gamma}}_{\mathrm{max}}(z)$} ;
\draw[draw=black,solid, -stealth, ->] (-1, -0.5) -- (-1, 1.5) node[above,right] {$z$};
\draw[draw=black,solid, -stealth, ->] (-1, -0.5) -- (1, -0.5) node[above,right] {$r$};
\end{tikzpicture}
\caption{Flow profile in the $r-z$ plane with different colors  indicating different velocities, with blue $0$ m$\,$s$^{-1}$ to red $0.007$ m$\,$s$^{-1}$. The shear band is the pink and light blue area, while the arrows indicate 10 $\%$ and 80 $\%$ cut-off range of shear rate as specified in the text.}
\label{theoretical}
\end{figure}

We obtain the macroscopic quantities by temporal averaging as explained in Sec. \ref{Averaging}. Next we analyze the data, neglecting data near walls ($r < r_\mathrm{min} \approx 0.045$ m, $r > r_\mathrm{max} \approx 0.105$ m, $z < z_\mathrm{min} \approx 0.004$ m) and free surface ($z > z_\mathrm{max} \approx 0.035$ m) as shown in Figure \ref{theoretical}. Further, the consistency of the local averaged quantities also depends on the local shear accumulated over time. We focus our attention in the region where the system can be considered to be in the critical state. This state is reached after large enough shear, when the materials deform with applied strain without any change in the local quantities, independent of the initial condition. To determine the region in which the flow is in critical state, $\dot{\gamma}_\mathrm{max}(z)$ is obtained from the maximum over all the strain rate at different $r$ for a given height $z$ or pressure. The critical state is achieved at a given pressure over regions with strain rate larger than the strain rate $0.1 \dot{\gamma}_\mathrm{max}(z)$ as shown in Figure \ref{theoretical} corresponding to the region of shear band. Most of our analysis shown in the latter sections are in the shear band center ($0.8 \dot{\gamma}_\mathrm{max}(z)$) at different heights in the system. However, we extend our studies to the shear-rate dependence in critical state which is effective for critical state data for wider regions of shear band (Sec. \ref{sec:nonlocal}). This shear rate dependence is analyzed in the regions of strain rate larger than the $0.1 \dot{\gamma}_\mathrm{max}(z)$ at a given height $z$. These data include the region from the center to the tail of the shear band, with typical cut-off factors $s_c = 0.8$ or $0.1$, respectively, as shown in Figure \ref{theoretical}, and explained in Sec. \ref{sec:nonlocal}.
\section{Time scales}\label{timescale}
Dimensional analysis is often used to define the characteristic time scales for different physical phenomena that the system involves. Even in a homogeneously deforming granular system, the deformation of individual grains is not homogeneous. Due to geometrical and local parametric constraints at grain scale, grains are not able to displace as affine continuum mechanics dictates they should. The flow or displacement of granular materials on the grain scale depends on the timescales for the local phenomena and interactions. Each time scale can be obtained by scaling the associated parameter with a combination of particle diameter $d_p$ and material density $\rho$. While some of the time scales are globally invariant, others are varying locally. The dynamics of the granular flow can be characterized based on different time scales depending on local and global variables. First, we define the time scale related to contact duration of particles which depends on the contact stiffness $k$ as given by \cite{singh2015role}:
\begin{equation} \label{Tscale_stiff}
t_{k} = \sqrt{\frac{\rho{d_p}^3}{k}}~.
\end{equation}
In the special case of a linear contact model, this is invariant and thus represents a global time scale too. Two other time scales are globally invariant, the cohesional time scale $t_c$ , i.e. the time required for a single particle to traverse a length scale of ${d_p}/2$ under the action of an attractive capillary force and the gravitational time scale $t_g$, i.e. the elapsed time for a single particle to fall through half its diameter $d_p$ under the  influence of the gravitational force. The time scale $t_c$ could vary locally depending on the local capillary force $f_c$. However, due to not taking into account liquid transport, and since the capillary force is weakly affected by the liquid bridge volume anyway while it strongly depends on the surface tension of the liquid $\sigma$, leaves the cohesion time scale as a global parameter given by:
\begin{equation} \label{Tscale_coh}
t_{c} = \sqrt{\frac{\rho{d_p}^4}{f_c}} \propto \sqrt{\frac{\rho{d_p}^3}{\sigma}}~,
\end{equation}
with surface tension $\sigma$ and capillary force $f_c \approx {\pi}\sigma{d_p}$.
The corresponding time scale due to gravity which is of significance under small confining stress close to the free surface is defined as:
\begin{equation} \label{Tscale_gravity}
t_{g} = \sqrt{\frac{d_p}{g}}~.
\end{equation}
The global time scales for granular flow are complemented by locally varying time scales. Granular materials subjected to strain undergo constant rearrangement and thus the contact network re-arranges (by extension and compression and by rotation) with a shear rate time scale related to the local strain rate field:
\begin{equation} \label{Tscale_strainrate}
t_{\dot{\gamma}} = {\frac{1}{{\dot{\gamma}}}}~.
\end{equation}
Finally, the time for rearrangement of the particles under a certain pressure constraint is driven by the local pressure p. This microscopic local time scale based on pressure is:
\begin{equation} \label{Tscale_pressure}
t_{p} = d_p\sqrt{\frac{\rho}{p}}~.
\end{equation}
As the shear cell has an unconfined top surface, where the pressure vanishes, this time scale varies locally from very low (at the base) to very high (at the surface). Likewise, the strain rate is high in the shear band and low outside, so that also this time scale varies between low and high, respectively.
\par
Dimensionless numbers in fluid and granular mechanics are a set of dimensionless quantities that have a dominant role in describing the flow behavior. These dimensionless numbers are often defined as the ratio of different time scales or forces, thus signifying the relative dominance of one phenomenon over another. In general, we expect five time scales ($t_g$, $t_p$, $t_c$, $t_{\dot{\gamma}}$ and $t_k$) to influence the rheology of our system. 
All the dimensionless numbers in our system are discussed in brief in the following two sections of this paper for the sake of completeness, even though not all are of equal significance.

\section{Rheology of dry granular materials}
 
\subsection{Effect of softness in the bulk of the materials}\label{sec:softness}
\begin{figure}[!htb]
  \begin{center}
  {%
	\includegraphics[width=0.6\columnwidth]{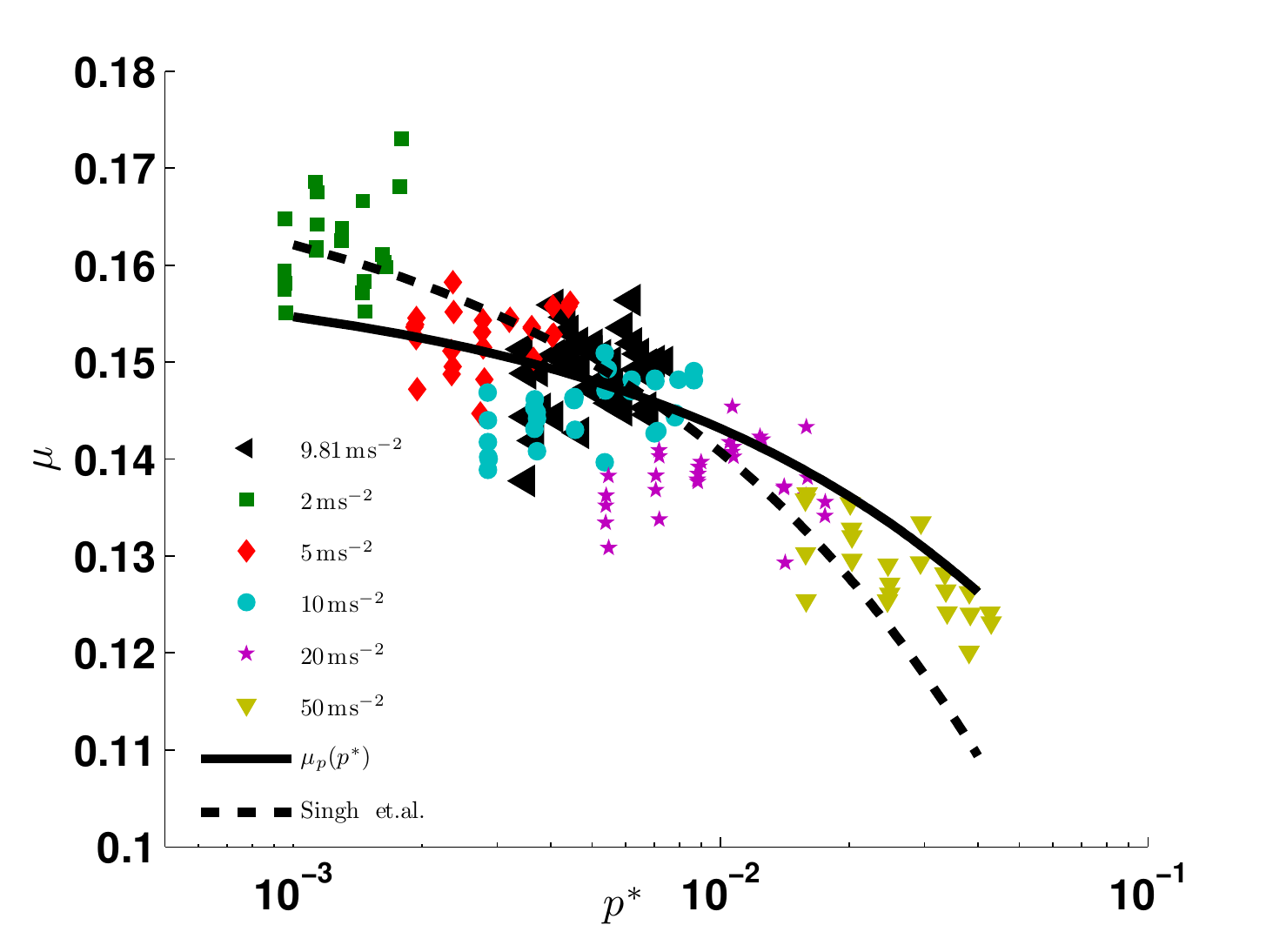}
   }%
  \end{center}
  \caption{Local friction coefficient $\mu$ as a function of softness $p^*$ for data with different gravity $g$ \cite{singh2015role} and our data (represented by $\blacktriangleleft$) for ${p_g}^* > 10$. The solid line represents the function ${{\mu}_{p}(p^*)}$.}
  \label{fig:softness}
\end{figure}

We study here the effect of softness on macroscopic friction coefficient for different gravity in the system. Thus the pressure proportional to gravity is scaled in dimensionless form $p^*$ \cite{singh2015role} given by:
\begin{equation} \label{effsoft}
{p}^* = \frac{pd_p}{k}~.
\end{equation}
This can be interpreted as the square of the ratio of time scales, $p^* = {{t_k}^2}/{{t_p}^2}$, related to contact duration and pressure respectively. Figure \ref{fig:softness} shows the macroscopic friction coefficient as a function of the dimensionless pressure $p^*$ and the 
dashed line is given by:
\begin{equation} \label{mu(I,P)}
{\mu}_p(p^*) = {{\mu}_o}{{f}_p(p^*)} \ \ \mathrm{with} \ \ {{f}_p(p^*)} = \bigg{[}1 - {({p^*}/{{p_o}^*})}^{\beta}\bigg{]}~,
\end{equation}
where, $p^* \ll {p_o}^*$ and ${p_o}^*$ denotes the limiting dimensionless pressure around the correction due to softness of the particles is not applicable anymore, since $f_p \leq 0$ for $p^* \geq {p_o}^*$, $\beta \approx 0.50$ and  ${p_o}^* \approx 0.90$ \cite{luding2016particles}. This is in close agreement with the plotted data. We compare our data with data for different gravity in the system from \cite{singh2015role}. The solid line represents the softness correction as proposed by \cite{singh2015role}. The effect of softness is dominant in regions of large pressure where the pressure time scale $t_p$ dominates over the stiffness time scale $t_k$ and thus the data in plot are corresponding to higher than a critical pressure (${p_g}^* > 10$, explained in Sec. \ref{sec:smallpr}). Here, the compressible forces dominate over the rolling and sliding forces on the particles, the flow being driven by squeeze. Thus, the macroscopic friction coefficient decreases with softness.

\begin{figure}[!htb]
  
\end{figure}
\subsection{Effect of inertial number}
\begin{figure}[!htb]
  \begin{center}
  {%
	\includegraphics[width=0.6\columnwidth]{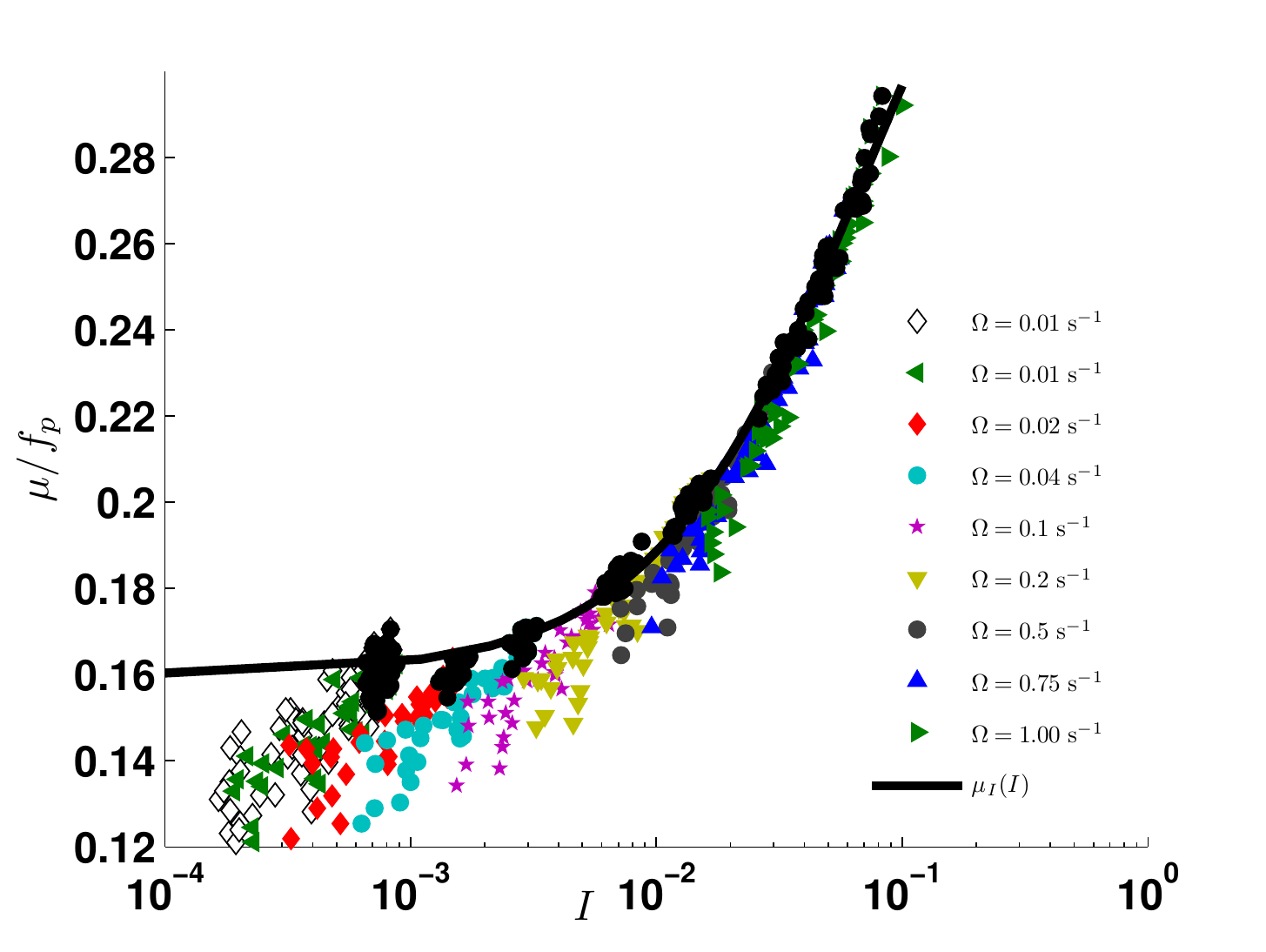}
   }%
  \end{center}
  \caption{Local friction coefficient $\mu$ scaled by the softness correction $f_p$ as a function of inertial number $I$. Different colors indicate different rotation rate $\Omega$ with our data represented by $\diamondsuit$. Black circles represent the data in the center of the shear band, other data are shown for $\dot{\gamma} > 0.1{{\dot{\gamma}}_{\mathrm{max}}}$ The solid line represents the function given by the ${\mu}_I(I)$ rheology given by Eq. (\ref{mu(I)}).}
  \label{fig:local}
\end{figure}

For granular flows, the rheology is commonly described by the  dimensionless inertial number \cite{midi2004dense}:
\begin{equation} \label{IN}
I = \dot{\gamma}d_{\mathrm{p}}/\sqrt{p/\rho} ~,
\end{equation}
which can be interpreted as the ratio of the time scales, $t_{p}$ for particles to rearrange under pressure $p$, and the shear rate time scale ${t_{\dot{\gamma}}}$ for deformation due to shear flow, see Sec. \ref{timescale}. It has been shown both experimentally \cite{forterre2008flows,jop2006constitutive,midi2004dense} and in simulations \cite{pouliquen2006flow} that for intermediate inertial numbers $0.01 < I < 0.5$, the macroscopic friction and the objective macroscopic friction both follow the so-called $\mu(I)$ rheology:
\begin{equation} \label{mu(I)}
{\mu}_I (I) = {\mu}_o + ({\mu}_{\infty} - {\mu}_o)\frac{1}{1+{I}_o/I}~,
\end{equation}
We assume the combined effect of softness and inertial number given as $\mu(p^*,I) = \mu_I(I)f_p$ and thus analyse $\mu/f_p$ as a function of $I$, see figure \ref{fig:local}. 
We compare our data for non-cohesive materials which is shown to be in agreement with the trend of data obtained from \cite{singh2015role} for different external rotation rates. The black solid line corresponds to the data in the shear band center ($\dot{\gamma} > 0.8 {\dot{\gamma}_\mathrm{max}}$ fitted by Eq. (\ref{rotr}) with ${\mu}_o = 0.16$, ${\mu}_{\infty} = 0.40$ and $I_o = 0.07$ which are in close agreement with the fitting constants explained in \cite{luding2016particles}. 
\subsection{Effect of gravity close to the free surface}\label{sec:smallpr}
\begin{figure}[!htb]
  \begin{center}
  {%
	\includegraphics[width=0.6\columnwidth]{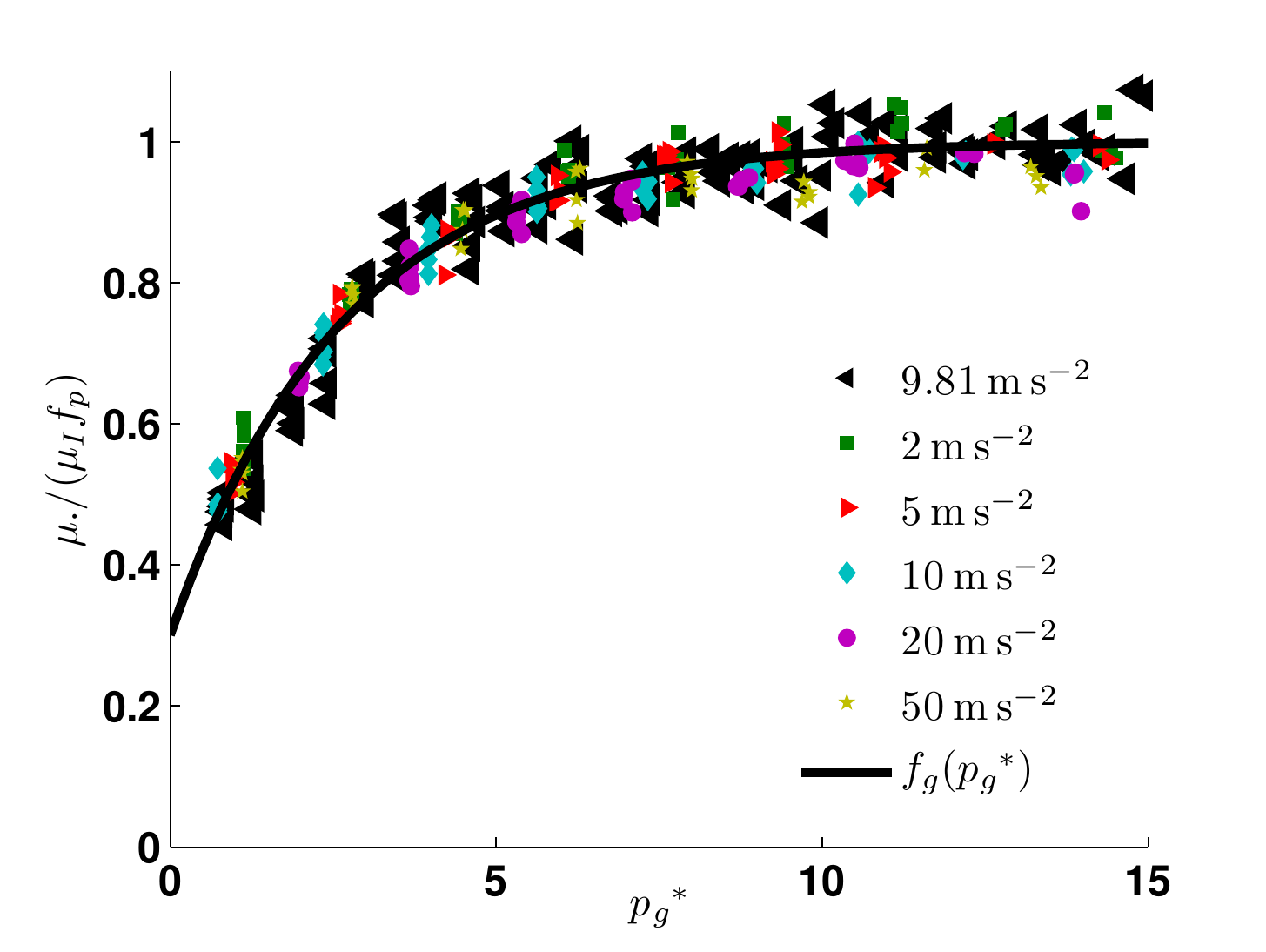}
   }%
  \end{center}
  \caption{Local friction coefficient $\mu$ scaled by softness correction ${f}_p$ and the inertial number correction ${\mu}_I$ as a function of dimensionless pressure ${p_g}^*$ for data with different gravity $g$. Different symbols and colors indicate different $g$ with our data represented by $\blacktriangleleft$. The solid line represents the function $f_g({p_g}^*)$.}
  \label{smallp}
\end{figure}

In this section, we investigate the effect of the another dimensionless number ${p_g}^*$ on local
friction coefficient, given by:
\begin{equation} \label{eq:gravity}
{p_g}^* = \frac{p}{\rho{d_\mathrm{p}}g}~.
\end{equation}
This can be interpreted as the square of the ratio of time scales, ${p_g}^* = {{t_g}^2}/{{t_p}^2}$, related to gravity and pressure respectively. The effect of inertial number and softness correction are eliminated by scaling $\mu$ by the correction factors ${\mu}_I$ and ${f}_p$ respectively and studying the effect of ${p_g}^*$ on the scaled friction coefficient. Figure \ref{smallp} shows $\mu$
scaled by ${\mu}_p$ as a function of dimensionless pressure ${p_g}^*$ for different gravity $g$ including our data for $g = 9.81$ ms$^{-2}$ which is also in agreement with other data set. All the data for different gravity collapse and these can be fitted by the solid line given by the correction $f_g({p_g}^*)$ where:
\begin{equation} \label{eq:smallp}
{\mu}_g({p_g}^*) = \mu_o{f}_g({p_g}^*)\ \  \textnormal{with} \ \ {f}_g({p_g}^*) = \bigg{[}1- a'\exp{\bigg(-\frac{{p_g}^*}{{p_{go}}^*}\bigg)}\bigg{]}~,
\end{equation}
where, $a' \approx 0.71$ is the relative drop in friction coefficient at ${p_g}^* = 0$, ${p_{go}}^* \approx 2.56$ is the dimensionless pressure at which the friction coefficient drops below $0.74{\mu}_o$ and $f_g({p_g}^*)$ is the correction corresponding to the dimensionless pressure ${p_g}^*$. Due to lack of confining stress close to the free surface (${p_g}^* < 10$), the macroscopic friction coefficient exponentially decreases with decrease in ${p_g}^*$. Here, the gravity time scale $t_g$ dominates over the pressure time scale $t_p$. Thus, while the effect of gravity close to the free surface is dominant for ${p_g}^* < 10$, ${p_g}^* \approx 10$ is the critical pressure above which the effect of softness $p^*$ is significant as explained in Sec. \ref{sec:softness}.

\subsection{Shear rate dependence in critical state flow} \label{sec:nonlocal}
\begin{figure}[!htb]		
\begin{center}
  {%
	\includegraphics[width=0.6\columnwidth]{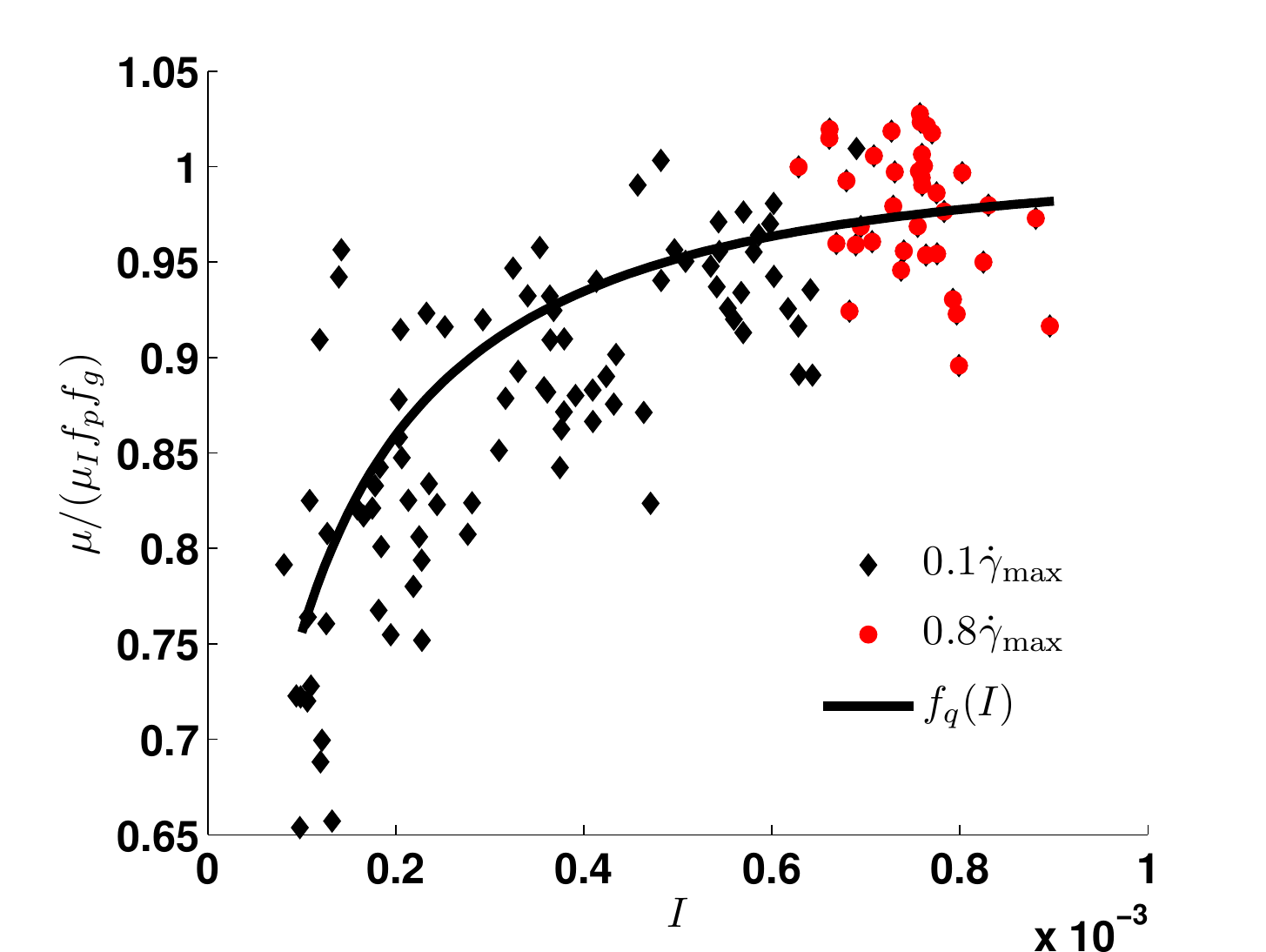}
   }%
  \end{center}
\caption{Local friction coefficient $\mu$ scaled by correction factors $f_p$, $f_g$ and ${\mu}_I$ as a function of inertial number $I$ for dry non-cohesive materials with data for $p^* > 0.003$. The solid line represents the function $f_q(I)$ from Eq. (\ref{mu(I,P)_small}).}
\label{fig:nonlocal}
\end{figure}

After having quantified the dependence of the macroscopic friction on inertial number and softness, another correction was proposed in \cite{singh2015role}, taking into account a reduced, relaxed friction correction in very slow quasi-static flow. The same phenomena was adddressed in \cite{kamrin2012nonlocal,koval2009annular,luding2008constitutive}  using non-local constitutive relations. Figure \ref{fig:nonlocal} is a representation of this correction $f_q(I)$ where:
\begin{equation} \label{mu(I,P)_small}
{\mu}_q(I) = {{\mu}_o}{f}_q(I) \ \ \textnormal{with} \ \  {f}_q(I) = \bigg{[}1 - \exp{\bigg(-{\bigg(\frac{I}{I^*}\bigg)}^{{\alpha}_1}\bigg)}\bigg{]}~,
\end{equation}
where, $I^* = (4.85\pm1.08)\times 10^{-5}$ for very small inertial numbers ($I \leq I^* $) and ${\alpha}_1 = 0.48\pm0.07$. $I^*$ scales linearly with the external shear rate and thus is proportional to the local strain-rate and the granular temperature \cite{singh2015role}. Although the data represented in figure \ref{fig:nonlocal} (black $\diamond$ and red $\circ$) include ${\dot{\gamma}}_c(z) > 0.1{\dot{\gamma}}_{\mathrm{max}}(z)$, the fitted solid line given by ${f}_q(I)$ correction corresponds to data in the shear band center as well as outside center (for ${\dot{\gamma}}_c(z) > 0.1{\dot{\gamma}}_{\mathrm{max}}(z)$) which are all in the critical state. Typically, we study the local effect for data inside the shear band center (${\dot{\gamma}}_c(z) > 0.8{\dot{\gamma}}_{\mathrm{max}}(z)$) which corresponds to the data given by red $\circ$ which are invariant to the effect of small inertial number which allows us to assume $f_q(I) \approx 1.0$. Hence, in the
following sections, we do not take into consideration the correction $f_q(I)$, though we mention it.

\section{Rheology of wet-cohesive granular materials}

\subsection{Bond number}
 The Bond number (${Bo}$) is a measure of the strength of the adhesive force relative to the compressive force. A low value of ${Bo}$ (typically much less than $1$) indicates that the system is relatively unaffected by the attractive forces; high  $Bo$ indicates that the attractive force dominates in the system. Thus $Bo$ is a critical microscopic parameter that controls the macroscopic local rheology of the system. We introduce here the local (simulation based) Bond number as:  
\begin{equation} \label{Bo_l}
{Bo}\left(p\right) = \frac{{f_c}^{\mathrm{max}}}{p{d_p}^2}~,
\end{equation}
defined as the square of the ratio between timescales related to pressure $t_{p}$ and wetting time scale $t_{c}$. ${f_c}^{\mathrm{max}} = 2\pi{r}\gamma\cos{\theta}$ is the maximum capillary force between a pair of particles, where $r$ is the effective radius of the interacting pair of particles. This provides an estimate of the local cohesion intensity by comparing the maximum capillary pressure allowed by the contact model ${{f_c}^{\mathrm{max}}}/{{d_p}^2} $ with the local pressure. A low to high transition of local Bond number from the bottom of the shear cell to the free surface is as a result of the change in time scale related to pressure $t_{p}$ from $t_{p} \ll t_{c}$ to  $t_{p} \gg t_{c}$ respectively. Subsequently, we define the global Bond number ${Bo}_g$ as a measure of the strength of cohesion in the system as:
\begin{equation} \label{Bo_g}
{Bo}_g = \frac{{f_c}^{\mathrm{max}}}{p~^{\mathrm{mean}}{d_p}^2}~,
\end{equation}
where, $p^{\mathrm{mean}}$ is the mean pressure in the system. This is an experimentally measurable quantity and is related to quantifying the system as a whole. The global Bond number corresponding to surface tension of liquid defined in Eq. (\ref{Surf}) is given by:
\begin{equation} \label{Bol}
{Bo}_g \ \in \ \left\{0.0\right., \ 0.06,  \  0.12, \ 0.24, \
 0.36,  \ 0.60,  \   1.28,   \   1.94,  \  2.54, \ \left. 3.46\right\} \  
\end{equation}
 
\subsubsection{Effect of local Bond number}

\begin{figure}[!htb]
  \begin{center}
  {%
	\includegraphics[width=0.6\columnwidth]{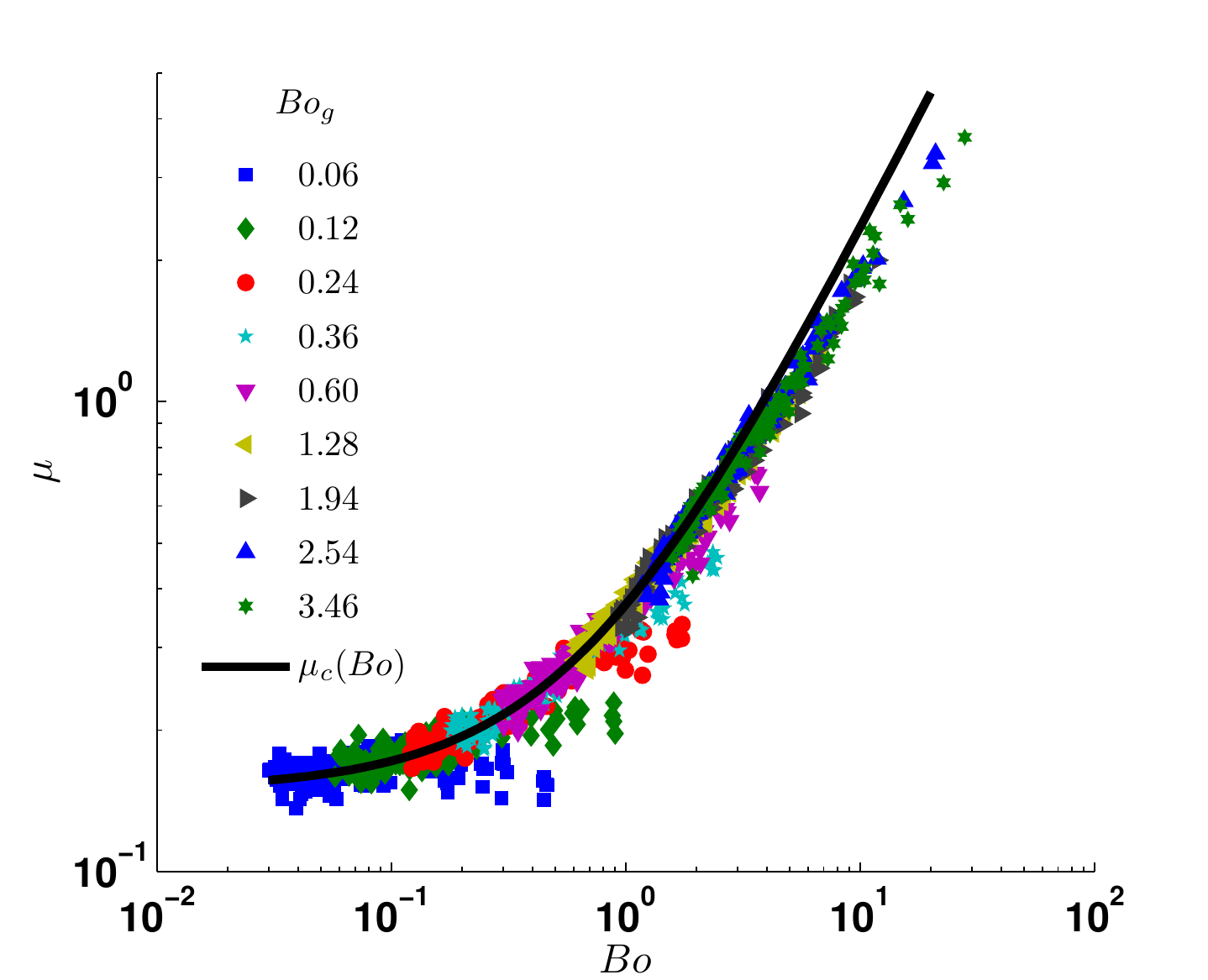}
   }%
  \end{center}
  \caption{Local friction coefficient $\mu$ as a function of the local Bond number ${Bo}$ for wet cohesive materials. The solid line is given by Eq. (\ref{mu}).}
  \label{mu_lim}
\end{figure}

The properties of the particles and the interstitial fluid strongly affect the macroscopic behavior of granular materials. The local macroscopic friction is studied as a function of local Bond number ${Bo}$ for different wet cohesion intensity. 
Figure \ref{mu_lim} shows the macroscopic friction coefficient as a function of the local Bond number ${Bo}$ for different wet cohesion. It is evident that the friction coefficient increases with local Bond number with a constant value ${\mu}_{o}$ in the low Bond number limit. For frictionless wet cohesive materials, the rheology can be defined by a linear fitting function given by:
\begin{equation} \label{mu}
{\mu}_c(Bo) = {{\mu}_{o}}f_c(Bo) \ \ \textnormal{with} \ \ f_c(Bo) = (1 + a{Bo}) ~,
\end{equation}
where, ${\mu}_{o} = 0.15$ is the macroscopic friction coefficient in the high pressure limit \cite{roy2015microtextendashmacro} and $a \approx 1.47$. This is shown by the solid line in figure \ref{mu_lim}. However, it is observed that the data deviate from the solid fitting line in the high Bond number or low pressure limit. This deviation is explained by the small pressure correction $f_g({p_g}^*)$ as explained in Sec. \ref{sec:smallpr} and discussed in details in the next section.

\subsection{Effect of gravity for wet materials}
\begin{figure}[!htb]
  \begin{center}
  {%
	\includegraphics[width=0.6\columnwidth]{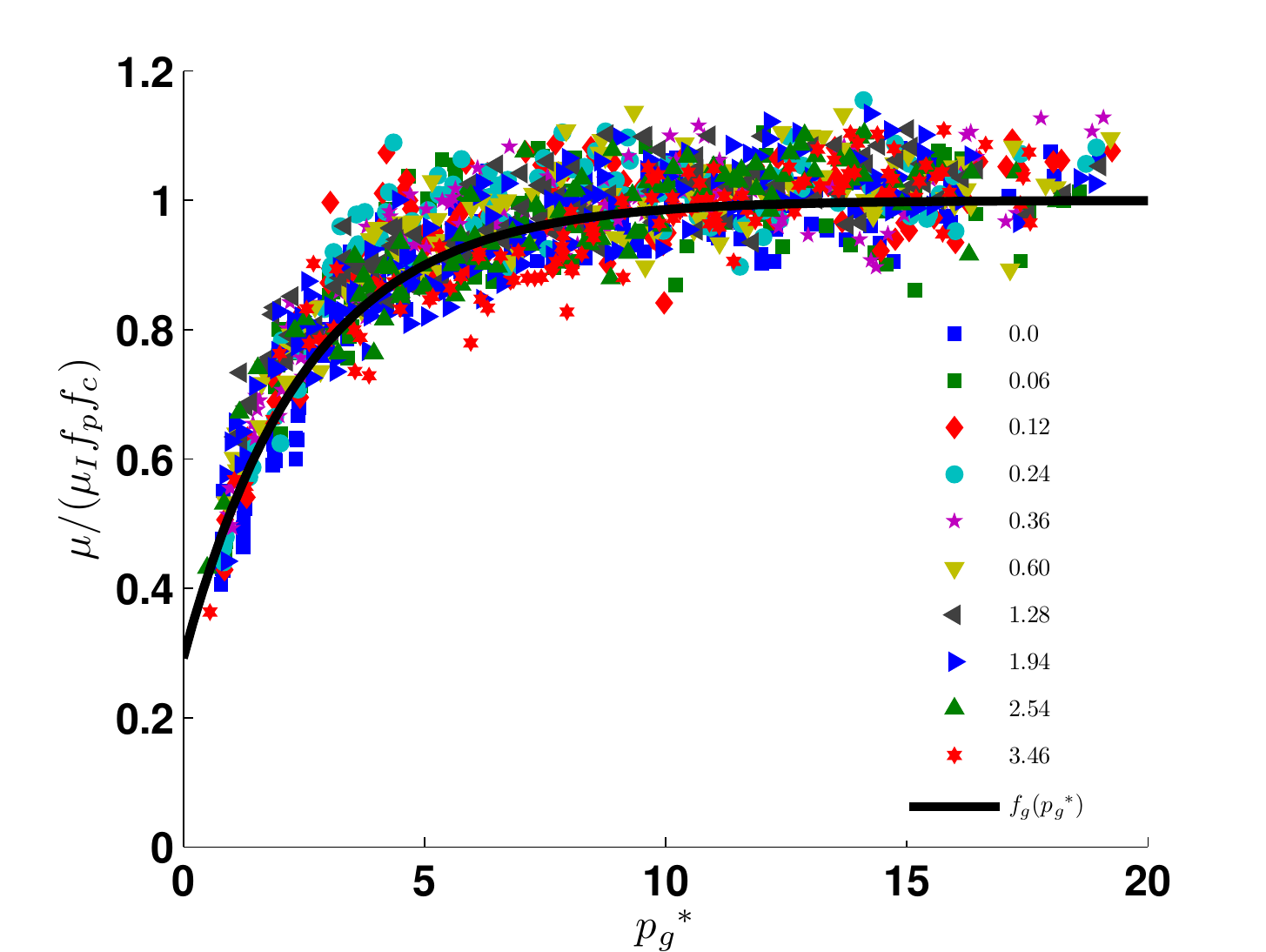}
   }%
  \end{center}
  \caption{$\mu/{({\mu}_I{f}_p{f}_c)}$ as a function of dimensionless pressure ${p_g}^*$ for different global Bond number. The solid line represents the function given by Eq. (\ref{eq:smallp}).}
  \label{smallprcoh}
\end{figure}

Figure \ref{smallp} shows the dependence of the local friction coefficient on the local scaled pressure ${p_g}^*$ for dry non-cohesive materials and this effect is small in the high pressure limit. With an attempt to separate the effect of Bond number on the rheology of cohesive materials, we plot the local friction coefficient $\mu$ scaled by the Bond number correction ${f}_c$ and other corrections $\mu_I$ and $f_p$, as a function of scaled pressure ${p_g}^*$ as shown in Figure \ref{smallprcoh}. The solid line is given by Eq. (\ref{eq:smallp}). The solid line for fitting with non-cohesive system fits for the wet data as well.

\section{Rheological model}

We studied the rheology of dry and wet granular materials in terms of different dimensionless numbers and the trends are combined and shown to collectively contribute to the rheology as multiplicative functions given by: 

\begin{equation} \label{comprheo}
{\mu}(I,p^*,{p_g}^*, {Bo}) = {{\mu}_I(I)}f_{g}({p_g}^*)f_{q}(I)f_{c}({Bo})f_{p}(p^*)~.
\end{equation}

Thus, a complete multiplicative function is obtained for the macroscopic friction coefficient as a dependence on four dimensionless numbers $p^*,~{p_g}^*, ~I,~  {Bo}$. \ref{appendixa} gives the summary and details of our proposed rheological model. 
\par

We have also analysed the volume fraction (not shown here) and it is observed to
decrease towards the surface and the shear band center. Note that the $I$ correction must not be neglected close to the free surface where $I$ can be largely increased. In the next section we analyse the effect of cohesion on the apparent viscosity and use the above model to predict it.

\section{Local apparent viscosity}

 To see the combined effect of pressure and strain rate on the local apparent viscosity, we analyse them as functions of the inertial number (since density if almost consistent). For a given pressure, the inertial number is proportional to the shear rate. Thus, the analysis of local apparent viscosity as a function of the inertial number for small pressure ranges can be interpreted as the analysis of viscosity vs strain rate. We define the dimensionless local viscosity as:
\begin{equation} \label{viscosity_nondim}
{\eta}^* = \frac{\eta}{\sqrt{d_\mathrm{p}k\rho}} = \frac{{\mu}p/\dot{\gamma}}{\sqrt{d_pk\rho}} = \frac{{\mu}p^*}{I}
\end{equation}
\par
Since we here focus on the data in the center of the shear band, the dependence on shear rate in the critical state flow which includes data outside the shear band center can be neglected ($f_q(I \ll 0) \approx 1$) and thus the rheological model for the local friction coefficient given by Eq. (\ref{comprheo}) is simplified by:
\begin{equation} \label{muI,p,{Bol}}
{\mu}({p}^*,{p_g}^*,{Bo}) = {{\mu}_I}(I)f_{g}({p_g}^*)f_{c}({Bo})f_{p}(p^*)~.
\end{equation}
The dimensionless variable ${\eta}^*$ can be related to three time scales namely, contact duration $t_{k}$, strain rate related time scale $t_{\dot{\gamma}}$ and pressure related time scale  $t_{{p}}$ as ${\eta}^* = \mu{t_{\dot{\gamma}}}t_{k}/{{t_{p}}^2}$.

\par
 Alternatively, the flow rules of granular materials can be approximated as that of a power-law fluid with inertial number inversely proportional to shear rate as given by:
\begin{equation} \label{flowbehavior}
{\eta}^*= KI^{\alpha - 1},
\end{equation}
where, $K = \mu{p^*}{I^{-\alpha}}$ is the flow consistency and $\alpha$ is the flow behavior index. The flow rules of granular materials are pretty straightforward at high pressures with $\alpha \approx 0$. However, deviations are observed from the power-law behavior at small pressures. More details on the flow rules at large and small pressure are explained in Sec. \ref{sec:largepr} and \ref{sec:small1pr} respectively.
\par
Figure \ref{fig:rheology} shows the local apparent viscosity ${\eta}^*$ as a function of the inertial number $I$ for different global Bond numbers. The data shown correspond to all the data close to the shear band center for different heights. The inertial number is lowest at an intermediate height, and increases towards surface and base. With increasing inertial number, the apparent shear viscosity decreases, indicating that granular materials flow like non-Newtonian fluids, specifically shear-thinning fluids. It is also evident from the figure that the flow behavior is different at large and small confining pressure.

\subsection{Prediction of local apparent viscosity}
\subsubsection{Prediction of strain rate}
Various numerical and experimental results suggest the presence of shear bands in granular materials subjected to relative motion \cite{fenistein2003kinematics,henann2013predictive}. Often this shear band is considered as a thin layer of localized strain rate, separating rigid blocks of constant velocity. Investigations on the shear band formation reveal that its characteristics are influenced by a number of factors including density, confining pressure, particle size and shape, friction, anisotropy of the material and cohesion \cite{henann2013predictive,singh2014effect}. The shear band thickness and the distance from the center decrease as the confining pressure increases \cite{ad1987values}. Constitutive relations exist for many shear band properties \cite{ries07}, which suggests a pathway to finding
analytical solutions.
\par
In this section, we discuss an analytical approach to get stress and strain rate correlations from the physics of granular materials and compare our analytical solution with the numerical results for different wet cohesion using the generalized $\mu$ function for the macroscopic friction, see Eq. (\ref{comprheo}) and (\ref{muI,p,{Bol}}). The magnitude of the strain rate is given by Eq. (\ref{s_r}).
It is assumed that the velocity component $u_{\phi}$ is slowly varying in \textit{z}-direction (${\partial u_{\phi}}/{\partial z} \approx 13\%$ of $({\partial u_{\phi}}/{\partial r} - u_{\phi}/{r})$ in the shear band center), so ${\partial u_{\phi}}/{\partial z}$ is small (by one order of magnitude) and is neglected with an approximation, so that
\begin{equation} \label{sr_s}
\dot{\gamma} \approx \frac{1}{2}\bigg{(}\frac{\partial{u_{\phi}}}{\partial{r}} - \frac{u_{\phi}}{r}\bigg{)}~.
\end{equation}
In the shear band region, the non-dimensionalized angular velocity profile $\omega = {u}_{\phi}/({2{\pi}r\Omega})$ at every height can be well approximated by an error function \cite{cohen2013flow,dijksman2010granular,fenistein2004universal,luding2008cohesive}:
\begin{equation} \label{fit}
\omega = A + B{\mathrm{erf}\left(\frac{r - {R_c}}{W}\right)}~,
\end{equation}
where $A \approx B \approx 0.5$, $W$ and $R_c$ are the width and the position of the shear band, respectively at different heights. Most surprising is the fact that the fit works equally well for a wide range of $I,~ p^*,~ Bo$ etc. \cite{singh2014effect}. Eq. (\ref{fit})  substituted in Eq. (\ref{sr_s}) can be simplified as a first order expansion of the derivative of the error function as: 
\begin{equation} \label{s_r1}
{\dot{\gamma}} = \frac{\sqrt{\pi}r{\Omega}}{W}\exp{\bigg[-{\bigg[\frac{{r - R_c}}{W}\bigg]}^2\bigg]}~.
\end{equation}
The shear rate at the center of the shear band ($r = R_c$) is thus given as:
\begin{equation} \label{s_r_max}
{\dot{\gamma}_\mathrm{max}} = \frac{{{\sqrt{\pi}}}R_c{\Omega}}{W}~.
\end{equation}
The pressure for the given geometry is increasing linearly from the free surface, \textit{i.e.} varies hydrostatically with the depth inside the material. Further, we obtain the non-dimensional inertial number from the predicted strain rate and pressure, so that
\begin{equation} \label{Imax}
{I_{\mathrm{max}}} = \frac{\dot{{\gamma}}_{\mathrm{max}}d_\mathrm{p}}{\sqrt{p/{\rho}_{\mathrm{max}}}} = \frac{\dot{{\gamma}}_{\mathrm{max}}d_\mathrm{p}}{\sqrt{H-z}}~,
\end{equation}
with ${\rho}_{\mathrm{max}} = \rho$, ignoring the small variations in density.
\subsubsection{Prediction of viscosity of materials confined to large pressure}\label{sec:largepr}
The predicted local apparent viscosity can be simplified with $f_g({p_g}^*) \approx 1$ under large pressure, ${\mu}_I(I) \approx \mu_o$ for quasistatic states and $f_p({p}^*) \approx 1$ for the relatively stiff
particles ($0.002 < p^* < 0.01$) studied in our system from Eqs. (\ref{viscosity_nondim}) and (\ref{muI,p,{Bol}}) and thus can be written as:
\begin{equation} \label{viscosity_high}
{\eta}^* = \frac{{\mu}_of_c(Bo)}{I}\sqrt{\frac{d_\mathrm{p}p}{k}} = \frac{{\mu}_o\sqrt{p^*}}{I}f_c(Bo)~.
\end{equation}

For dry non-cohesive materials, $f_c({Bo}) = 1$ and $\sqrt{p}$ is slowly changing at high pressure. For wet cohesive materials, the magnitude of viscosity is determined by the additional term $f_c({Bo})$. However, the flow behavior index for wet materials is also
constant under high confining pressure for the same reason as stated for the dry
materials. Table \ref{visc} shows the value of fitting coefficient $\alpha - 1$ for different ${Bo}_g$. Under high confined pressure, the flow behavior index $\alpha$ is independent of cohesion and
$\alpha \approx 0$ as shown in Table \ref{visc} and $\alpha - 1$ corresponding to the slope of the red dash-dotted lines in Figure \ref{fig:rheology}. Thus ${\eta}^* \propto I^{-1}$ and $\alpha \approx 0$ confirms that both dry and wet granular materials behave like a power law fluid under large confining pressure. 
\begin{table}[!htb]
\center
\caption{\label{visc}Table showing the flow behavior index under large pressure constraint}
\footnotesize
\begin{tabular}{@{}ccccccccccc}
\br

${Bo}_g$ & $0.0$ & $0.06$ & $0.12$ & $0.24$ & $0.36$ & $0.60$ & $1.28$ & $1.94$ & $2.54$ & $3.46$ \\
\mr
${\alpha} - 1$ & $-0.94$ & $-0.81$ & $-0.92$ & $-0.82$ & $-0.89$ & $-1.00$ & $-0.93$ & $-1.10$ & $-1.23$ & $-1.09$ \\ 
\br
\end{tabular}\\
\end{table}

\par

\subsubsection{Prediction of viscosity of materials confined to small pressure}\label{sec:small1pr}
\begin{figure}[!htb]

\centering
\includegraphics[width=.5\textwidth]{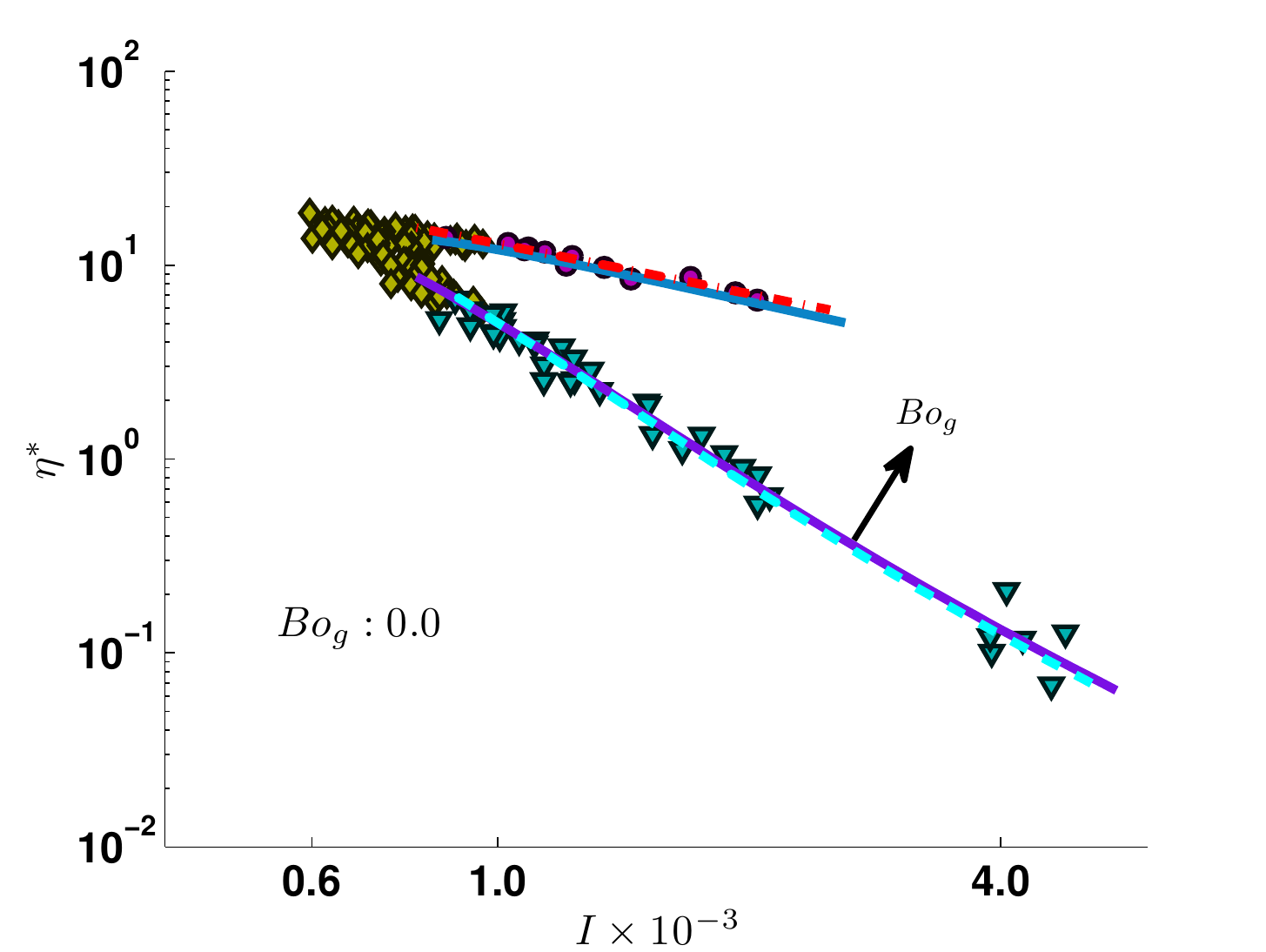}\hfill
\includegraphics[width=.5\textwidth]{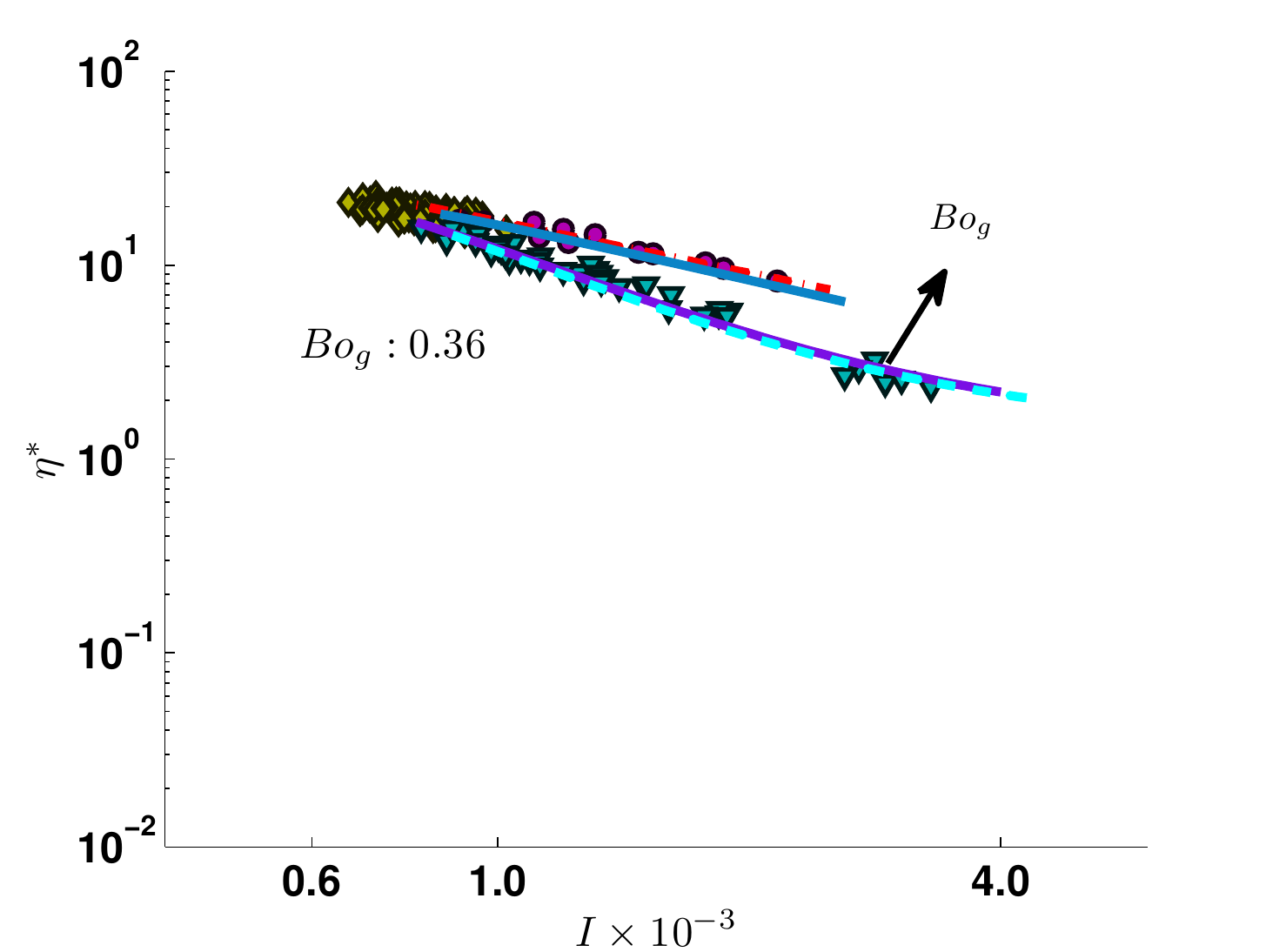}\\
\includegraphics[width=.5\textwidth]{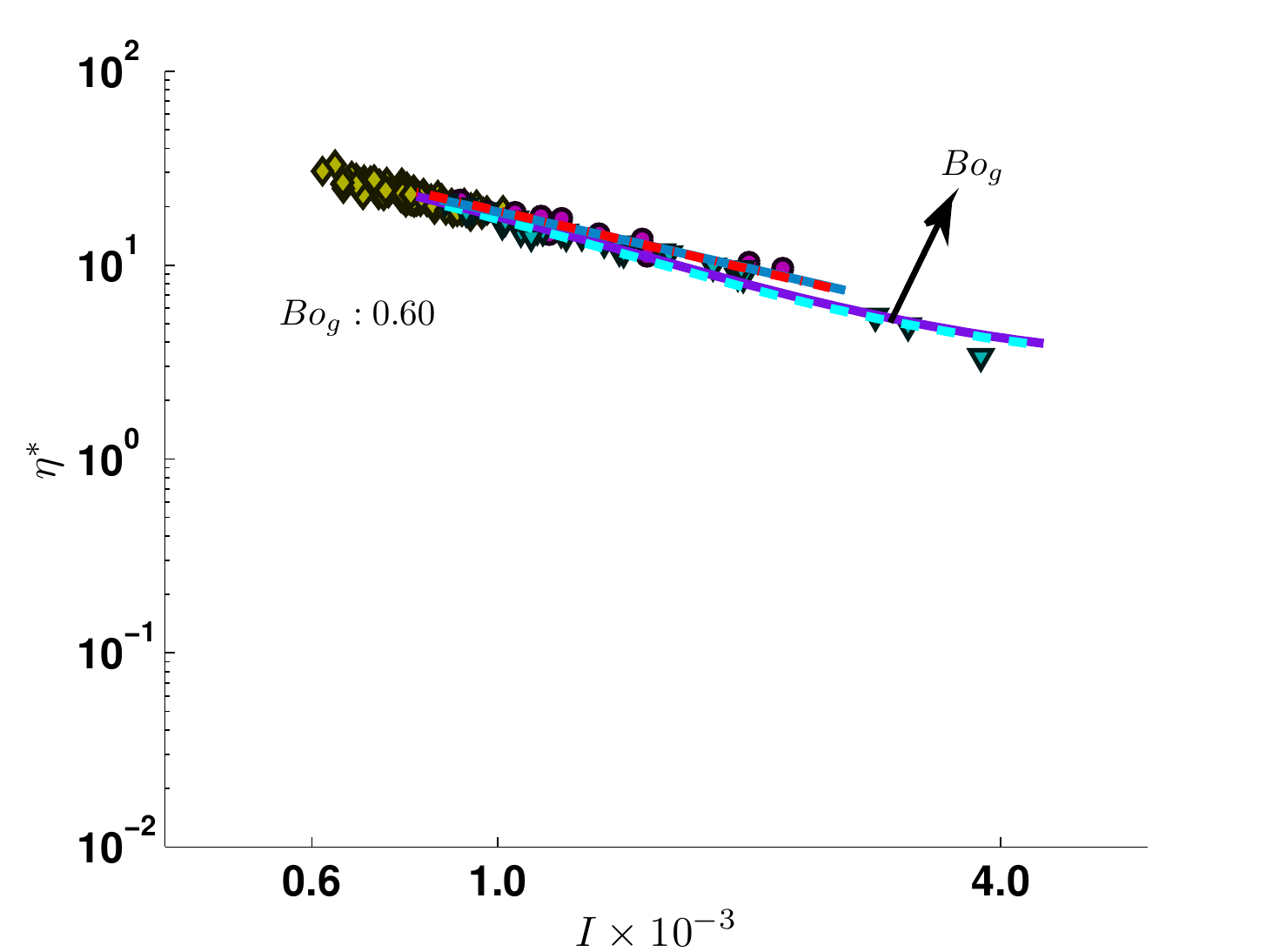}\hfill
\includegraphics[width=.5\textwidth]{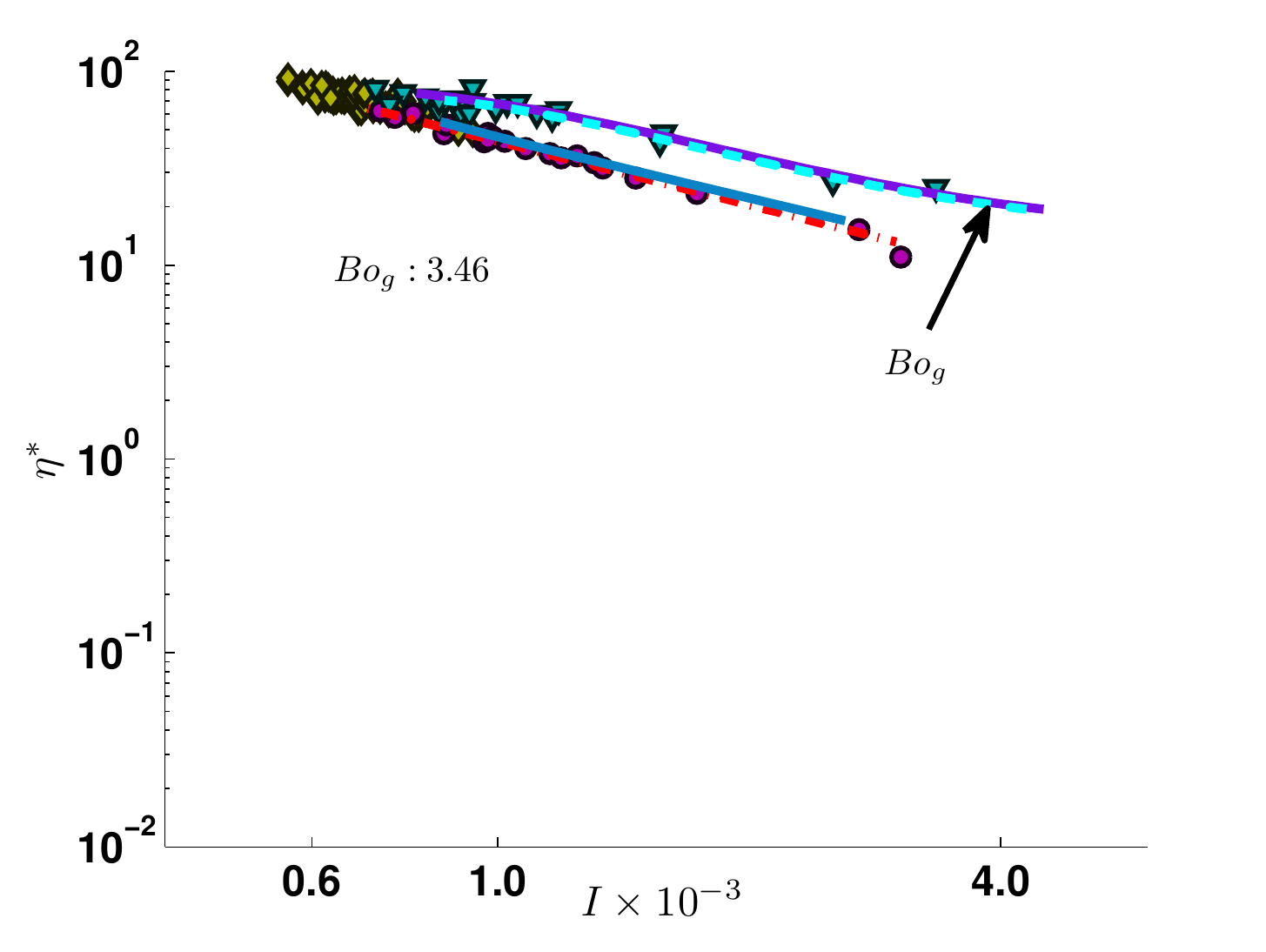}\\

\caption{Local apparent shear viscosity ${\eta}^*$ as a function of inertial number $I$ for different global Bond number ${Bo}_g$. Different symbols represent data for different pressure, $\bigtriangledown : p^*\,\geq 0.006\,$, $\diamond : 0.002\,<\,p^*\,<\,0.006\,$ and $\circ : p^*\,\leq\,0.002\,$. The fitting lines for $p^*\,\geq\,0.006\,$ (dash-dotted red) and $p^*\,\leq\,0.002\,$ (solid indigo) are given by Eqs. (\ref{flowbehavior}) and (\ref{viscosity_nondim}) respectively. The lines (solid blue) and (dashed cyan) are the predictions obtained from the analytical solution as explained in Sec. \ref{analytical} for $p^*\,\geq\,0.006\,$ and $p^*\,\leq\,0.002\,$ respectively.}
\label{fig:rheology}

\end{figure}

Wet cohesive materials confined to small pressure near the surface show more interesting behavior. Here, the pressure is very small, \textit{i.e.} large $t_p$ and $t_c$ make confining pressure and strain less dominant, so that $t_g$ and $t_c$ are the two interacting time scales. The rheology is now strongly dependent on the corrections $f_g({p_g}^*)$ and $f_c({Bo})$ and the correction $f_p({p}^*) \approx 1$ ($p^* < 0.005$) under small confining pressure. The strain rate close to the center and free surface is almost constant since the shear band is wide. We use this  simplified constant strain rate to predict the viscosity near the surface of the shear cell where the pressure is very small. The apparent shear viscosity for wet cohesive materials confined to small pressure is more intricate and is predicted by substituting Eqs. (\ref{viscosity}), (\ref{IN}) and (\ref{muI,p,{Bol}}) into Eq. (\ref{viscosity_nondim}), assuming a mean constant strain rate near the surface. Figure \ref{fig:rheology} shows the prediction of viscosity at small pressure as given by the indigo solid lines. Non-cohesive materials upto weakly cohesive materials (${Bo}_g < 0.60$), at low pressure, are less viscous than those at high pressure, as shown in the figure. For global Bond number ${Bo}_g = 0.60$, materials for a given inertial number have the same viscosity independent of pressure. For even higher cohesion (${Bo}_g > 0.60$), the flow behavior changes qualitatively. For a given inertial number, the material near the surface has higher viscosity than at the bulk and at the base. Materials confined by small pressure go more towards shear thickening with increase in cohesion. Thus, granular materials have different shear-thinning properties depending on the local confining pressure and Bond number. 

\subsubsection{Analytical prediction of viscosity}\label{analytical}
We extract the position and the width of the shear band $R_c$ and $W$ respectively from the fit function in Eq. (\ref{fit}). Both position and width of the shear band depend on the height in the system and the position moves inwards with increasing height (decreasing pressure). Predictions of the position of the shear band center as a function of height is given in \cite{unger2004force}. Since the analytical prediction discussed here is not significantly affected by this varying position of the shear band, we use the mean shear band position $\bar{R_{\mathrm{c}}}$ for our prediction. The shear band moves inward with increase in global Bond number \cite{singh2014effect}. Thus the mean shear band position $\bar{R_{\mathrm{c}}}$ decreases with increasing ${Bo}_\mathrm{g}$ (not shown here). 
\par
The width of the shear band is predicted as function of height as given by \cite{ries07}:
\begin{equation} \label{width2}
W\left(z\right) = W_{\mathrm{top}}{\left[1 - {\left(1 - \frac{z}{H}\right)}^2\right]} ^{\beta}~,
\end{equation}
where $\beta = 0.6$ for non-cohesive materials and $0.5 < \beta < 0.7$ for cohesive materials are fitted well by our data. Assuming the pressure varying hydrostatically and the bulk density as ${\rho} = 0.6{\rho}_{\mathrm{p}}$, we translate Eq. (\ref{width2}) to $W$ as a function of $p$. Substituting Eqs. (\ref{Imax}) and (\ref{width2}) in Eq. (\ref{IN}) and rearranging, we get the inertial number $I_\mathrm{max}$ in the shear band center as a function of the local pressure $p$. Further, by substituting $p$, we get ${\eta}^*_\mathrm{max}$ in the shear band center and thus obtain a quantitatively accurate prediction of ${\eta}^*_\mathrm{max}$ ($I_\mathrm{max}$), plotted as blue solid lines and cyan dashed lines in Figure \ref{fig:rheology}.
\par
The results show that the analytical solution is in good agreement with our numerical results. Focusing on the slope of the small pressure line, we observe that it changes with increasing cohesion in the same way as shown by numerical data. It is observed from the analytical solution that this change in slope is governed by  ${\mu}$. Thus, the shear-thinning rate for materials under small pressure depends on local friction coefficient, which depends on the corrections $f_g({p_g}^*)$ and $f_c(Bo)$.
 \subsection{Eliminating the effect of cohesion and gravity}\label{disincoh}
\begin{figure}[!htb]
		
	\includegraphics[width=0.5\columnwidth]{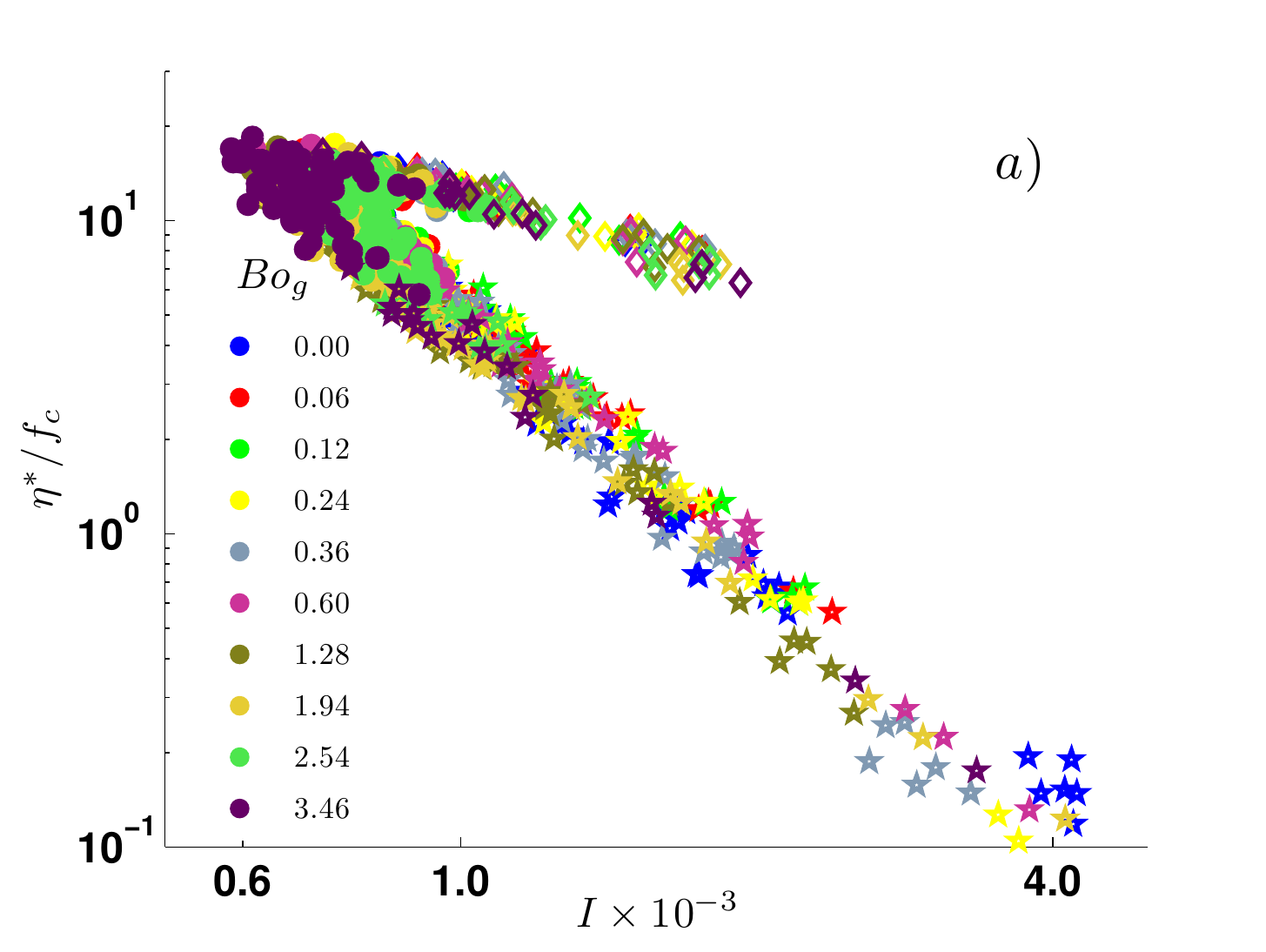}
\hfill
	\includegraphics[width=0.5\columnwidth]{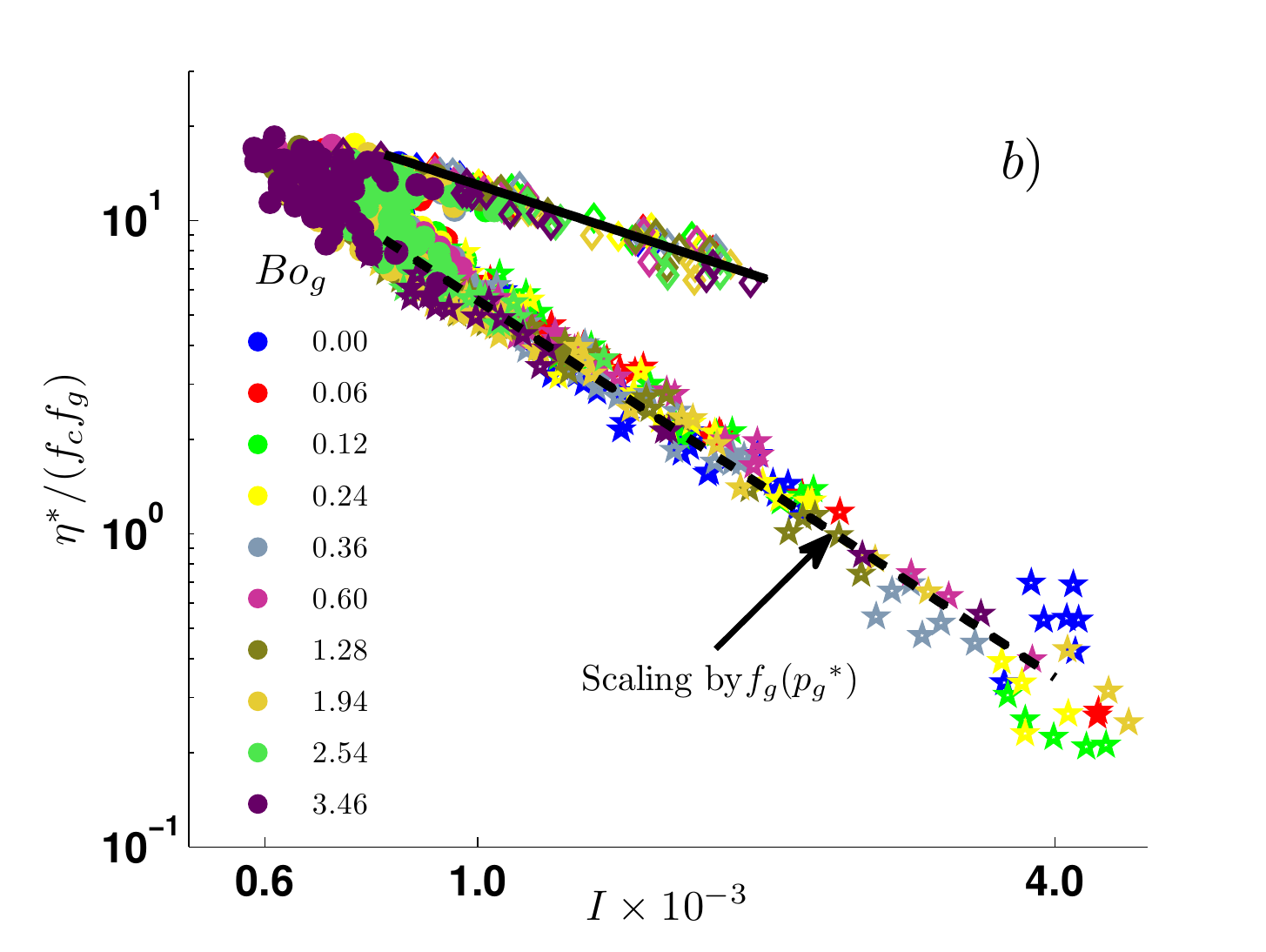}

\caption{a) Dimensionless viscosity ${\eta}^*$ scaled by the Bond number correction ${f}_c$ as a function of the inertial number $I$. b) Dimensionless viscosity ${\eta}^*$ scaled by the Bond number correction ${f}_c$ and small pressure correction ${f}_g$ as a function of the inertial number $I$. Different symbols represent data for different pressure, $\diamond\,:\,p^*\,\geq\,0.006\,$, $\bullet : 0.002\,<\,p^*\,<\,0.006\,$ and $\star\,:\,p^*\,\leq\,0.002\,$ respectively. The fitted solid and dashed lines for large and small pressure are given by Eq. (\ref{flowbehavior}) with $\alpha = 0$ and $\alpha = -1$ respectively.}
\label{fig:viscositycollapse}
\end{figure}

 Under larger confining pressure (as stated in Sec. \ref{sec:largepr}), with increase in cohesion, the viscosity of the granular fluid increases, however, the flow behavior remains qualitatively the same even for very high cohesion. For materials confined to large pressure, where $\sqrt{p}$ is slowly varying, the viscosity is inversely proportional to the strain rate and approximately also to the inertial number. At smaller pressure, the materials are more free only under the effect of gravity, with less dominant forces due to particle contacts. Therefore, cohesion is relatively more dominant for higher local Bond numbers, resulting in the qualitative change in shear thinning rate ($\alpha$). Thus the flow of materials is affected by both dimensionless numbers $Bo$ and ${p_g}^*$ at the same time. Then, the granular fluid appears to no longer behave like a power-law fluid. Several of these rheological correction factors make the flow behavior even more non-linear under small pressure. In order to see the rheology of the granular fluid under small pressure, which is devoid of the effect of these dimensionless numbers, we rescale the local dimensionless viscosity ${\eta}^*$ by ${f}_c(Bo)$ and $f_g({p_g}^*)$ and analyse it as a function of inertial number. Figure \ref{fig:viscositycollapse}a shows the dimensionless viscosity ${\eta}^*$ scaled by $f_c(Bo)$ as a function of inertial number for different cohesion. All the data for different cohesion collapse to a single plot for the triad of different pressure scales. Further, we rescale ${\eta}^*/{f_c(Bo)}$ by $f_g({p_g}^*)$ and plot it as a function of inertial number for different cohesion as shown in Figure \ref{fig:viscositycollapse}b. The fitted solid line corresponding to the data at large pressure is given by Eq. (\ref{flowbehavior}) with $\alpha = 0$ and $K \approx 0.01$. Furthermore, the fitted dashed line corresponding to the data at small pressure is given by Eq. (\ref{flowbehavior}) with $\alpha = -1$ and $K \approx 5.6\times{10}^{-6}$. This is explained theoretically by substituting Eq. (\ref{IN}) in Eq. (\ref{viscosity_nondim}) with constant friction coefficient ${\mu}_0$ yielding:
 \begin{equation} \label{viscosity_low}
{\eta}^* = \frac{{\mu}_{o}\dot{\gamma}{d_p}^{3/2}}{I^{2}}\sqrt{\frac{\rho}{k}}~,
\end{equation}
Thus, for slowly varying strain rate at small pressure, ${\eta}^*$ is proportional to $I^{-2}$ and is represented by Eq. (\ref{flowbehavior}) with $\alpha = -1$. This eventually explains the earlier observations by \cite{luding2008effect}.
\par
Thus, the flow behavior for granular materials in a simple hypothetical case with high confining stress constant friction coefficient can be approximated by that of a power-law fluid flow behavior. However, for more realistic systems, e.g., unit operations at low stress, several other factors influence the flow rheology, \textit{e.g.} near to the free surface. Thus, under small pressure, granular materials behave more interestingly and complex than a power-law fluid.
\section{Discussions and conclusions}
The rheology of dry as well as wet granular materials (in the pendular regime) has been studied by simulations using the discrete element method in steady state shear. It is observed that the local macroscopic friction coefficient varies significantly with cohesion and constraining pressure and thus design predictions must be improved accordingly. Our results show that the conventional $\mu(I)$ rheology must be modified to take into account other factors such as cohesion, contact softness, corrections at small pressures where gravity dominates, and a generalised inertial number dependence for very slow (creep) quasi-static flow. The trends are combined and shown to collectively contribute to the rheology as multiplicative functions, \textit{i.e.} ignoring one contribution can lead to inconsistent results. This new generalized rheological model applies to a wide range of parameters from dry non-cohesive to strongly cohesive materials, and contains also both the small and the large pressure limits. These results can be significant for industrial equipment design by predicting the handling difficulties due to joint effects of cohesion and confining pressure.
\par
Furthermore, we study the apparent viscosity as a function of inertial number for granular fluids of varying cohesive strength. Most strikingly, the cohesive strength not only increases the magnitude of the viscosity, but also decreases the shear thinning rate, but only for material under small confining pressure \textit{e.g.} close to the free surface. This variable shear thinning behavior of granular materials under low stress, close to a free surface, is attributed to the higher local Bond number and the low pressure effect. The flow rheology (friction and viscosity) is predicted by the proposed rheology models for dry as well as wet granular materials. Further, we develop an analytical solution for the apparent viscosity using the proposed rheology (with some simplifications) and show that the results are in good agreement with our numerical analysis. Materials become less and less shear thinning (or goes towards shear thickening) at high Bond numbers with an increase in cohesion under small confining pressure. This decrease in shear thinning behavior possibly can be explained by the similar phenomena as the hydrocluster formation in shear thickening fluids ``liquid body armor". Highly cohesive wet granular materials could thus become shear thickening. Then the systems no longer form cylindrical symmetric shear bands but clusters form and the study of such system is beyond the scope of this paper.  
\par
In the last section of this paper, we develop an analytical solution for predicting the viscosity of wet materials from the proposed rheology model. The analytical solution also agrees well with the numerical results and predicts the similar shear thinning rate under different confining pressure and hence this validates the new rheology model. Furthermore, it is shown that the effect of each of the dimensionless numbers can be eliminated by rescaling and thus the viscosity of a simple system with a (small) constant friction coefficient is predicted as that of a power-law fluid.

\par
We aim to implement the generalized rheological model to study the continuum description of a similar split-bottom shear cell geometry. A successful implementation is only the first step for validation and paves the way to use this rheological model in industrial applications for material flow descriptions. We aim to also include the higher order effect of the Bond number in the generalized rheology as an outlook. We include the small pressure correction in the rheology, stating it as an effect of gravity. It is to be noted that even in a micro-gravity system, both pressure and gravity changes identically and thus the correction corresponding to small pressure remains the same as in a system with gravity. Thus this correction corresponds to an effect of interface or free-surface.
\section*{Acknowledgements}

We would like to thank A. Singh as we could validate our simulations with his existing rheological dry data. We would also like to thank D. Berzi and D. Vescovi for their useful suggestions on our rheological models. This work is financially supported by STW Project 12272 ``Hydrodynamic theory of wet particle systems: Modeling, simulation and validation based on microscopic and macroscopic description".

\begin{appendix}
\section{Summary of the generalized rheological model}\label{appendixa}
\begin{table}[htb!]
\caption{List of rheological correction functions for application in a continuum model, see Eq. (\ref{comprheo})}
\begin{center}
\item[]\begin{tabular}{@{}l*{15}{l}l*{15}{l}l*{15}{l}}
\br
Dimensionless numbers&Corrections&Coefficients&Coefficients\cite{luding2016particles}\\
\mr
Inertial number ($I$)& $\mu_I = {\mu}_o + \frac{{\mu}_{\infty} - {\mu}_o}{1+{I}_o/I}$&\begin{tabular}{@{}c@{}c@{}}${\mu}_o = 0.16,\,$\\ ${\mu}_{\infty} = 0.40\pm0.01,\,$ \\ $I_o = 0.07\pm0.007$\end{tabular}&\begin{tabular}{@{}c@{}c@{}}${\mu}_o = 0.15,\,$\\ ${\mu}_{\infty} = 0.42,\,$ \\ $I_o = 0.06$\end{tabular} \\
\\
Softness ($p^*$)&$f_p = 1 - {({p^*}/{{p_o}^*})}^{\beta}$&Taken from \cite{luding2016particles} &\begin{tabular}{@{}c@{}c@{}}${\beta} = 0.50,\,$ \\ ${p_o}^* = 0.90$\end{tabular} \\
\\
Small pressure effect (${p_g}^*$)&$f_g = 1 - a'\exp{(-{{p_g}^*}/{{p_{go}}^*})}$&\begin{tabular}{@{}c}${a}' = 0.75\pm0.05,\,$ \\ ${p_{go}}^* = 2.30\pm0.30$\end{tabular} \\
\\
Nonlocal effect ($I$) &$f_q = 1 - \exp\bigg({-\bigg(\frac{I}{I^*}\bigg)}^{{\alpha}_1}\bigg)$&\begin{tabular}{@{}c@{}c@{}}${\alpha}_1 = 0.48\pm0.07,\,$ \\ ${I}^* = (4.85\pm1.08)\times10^{-5}$\end{tabular}& \begin{tabular}{@{}c@{}c@{}} \cite{luding2008constitutive} for a similar\\ correlation\end{tabular} \\
\\
Bond number ($Bo$) &$f_c = 1 + a{Bo}$&$a = 1.47\pm0.17$\\

\br
\end{tabular}
\end{center}
\end{table}
\end{appendix}
\section*{References}
\renewcommand{\section}[2]{}
\bibliographystyle{plain}       
\bibliography{library}   
\end{document}